\documentclass[aps,pra,12pt,nofootinbib,superscriptaddress,longbibliography,notitlepage]{revtex4-1}
\usepackage{eufrak}
\usepackage[T1]{fontenc} %for what? That's why: http://tex.stackexchange.com/
\usepackage{times}
\usepackage{amsthm} %for theorem baths
\usepackage{amssymb} %lots of math symbols
\usepackage{amsmath} %for spacing in equations and aligns
\usepackage{mathbbol} %for \1, load at last
\usepackage[normalem]{ulem}
\usepackage[usenames,dvipsnames]{color}
\usepackage{graphicx}
\usepackage{overpic}
\usepackage{xifthen}
\usepackage{verbatim}
\usepackage{booktabs}
\usepackage{float}
\usepackage{tasks}
\usepackage{paralist} %for compactenum

\definecolor{mydarkgreen}{rgb}{0,0.5,0}
\definecolor{mydarkred}{rgb}{0.5,0,0}
\definecolor{mydarkblue}{rgb}{0,0,0.5}

\usepackage{hyperref}
\hypersetup{
  colorlinks,
  citecolor=Blue,
  linkcolor=Blue,
  urlcolor=Blue}

\newcommand{\srcasea}{($\mathfrak{a}$)}
\newcommand{\srcaseb}{($\mathfrak{b}$)}
\newcommand{\srcasec}{($\mathfrak{c}$)}
\newcommand{\msrcasea}{(\mathfrak{a})}
\newcommand{\msrcaseb}{(\mathfrak{b})}
\newcommand{\msrcasec}{(\mathfrak{c})}
\newcommand{\msrcased}{(\mathfrak{d})}

\definecolor{c0}{HTML}{000000} % black
\definecolor{c1}{HTML}{9EC05B} % green
\definecolor{c2}{HTML}{3288BD} % blue
\definecolor{c3}{HTML}{D53E3F} % red
\definecolor{c4}{HTML}{E8B42B} % orange
\definecolor{c5}{HTML}{B82EE6} % purple
\definecolor{c9}{HTML}{FFFFFF} % white

\newcommand{\Tens}[4][]{\ifthenelse{\equal{#1}{}}
   {  % no optional argument
      #2_{#3}^{#4}
   }
   {  % optional argument
      #2_{#3}^{\lbrack #1 \rbrack \, #4}
   }
}

\makeatletter
\newcommand{\pushright}[1]{\ifmeasuring@#1\else\omit\hfill$\displaystyle#1$\fi\ignorespaces}
\makeatother

\newcommand{\1}{\mathbb{I}}

\newcommand{\ket}[1]{\left|{#1}\right\rangle}
\newcommand{\Lket}[1]{\left|{#1}\right\rangle\!\rangle}

\newcommand{\bra}[1]{\left\langle{#1}\right|}

\newcommand{\braket}[2]{\left\langle #1 \middle| #2 \right\rangle}

\newcommand{\statem}[1]{\ket{\uparrow \uparrow \uparrow}_{#1}}

\newcommand{\stateo}[1]{\ket{\uparrow \downarrow \uparrow}_{#1}}

\newcommand{\stater}[1]{\ket{\downarrow \uparrow \downarrow}_{#1}}

\newcommand{\statet}[1]{\ket{\downarrow \downarrow \downarrow}_{#1}}
\newcommand{\stateu}[1]{\ket{\uparrow \uparrow}_{#1}}
\newcommand{\statev}[1]{\ket{\downarrow \uparrow}_{#1}}
\newcommand{\statew}[1]{\ket{\uparrow \downarrow}_{#1}}
\newcommand{\statex}[1]{\ket{\downarrow \downarrow}_{#1}}
\newcommand{\statey}[1]{\ket{\uparrow}_{#1}}
\newcommand{\statez}[1]{\ket{\downarrow}_{#1}}

\renewcommand{\vec}[1]{\mathbf{#1}}

\definecolor{daniel}{rgb}{0,.1,1}

\definecolor{kenji}{RGB}{150,00,00} %random color

\definecolor{lincoln}{RGB}{40,180,40}

\newcommand{\csm}{Department of Physics, Colorado School of Mines, Golden,
  Colorado 80401, USA}
\newcommand{\lmu}{Department of Physics and Arnold Sommerfeld Center for
  Theoretical Physics, Ludwig-Maximilians-University Munich, 80333 Munich,
  Germany}

\begin{document}

\title{Thermalization in the Quantum Ising Model $-$ Approximations, Limits,
  and Beyond}

\author{Daniel Jaschke}
\affiliation{\csm}
\author{Lincoln D.\ Carr}
\affiliation{\csm}
\author{In\'es de Vega}
\affiliation{\lmu}

% =============================================================================
% Abstract
% =============================================================================

\begin{abstract}

We present quantitative predictions for quantum simulator experiments on
Ising models from trapped ions to Rydberg chains and show how the
thermalization, and thus decoherence times, can be controlled by considering
common, independent, and end-cap couplings to the bath. We
find (i) independent baths enable more rapid thermalization in
comparison to a common one; (ii) the thermalization timescale
depends strongly on the position in the Ising phase diagram;
(iii) for a common bath larger system sizes show a significant
slow down in the thermalization process; and (iv) finite-size
scaling indicates a subradiance effect slowing thermalization rates
toward the infinite spin chain limit. We find it is necessary
to treat the full multi-channel Lindblad master equation rather
than the commonly used single-channel local Lindblad approximation
to make accurate predictions on a classical computer. This method reduces
the number of qubits one can practically classical simulate
by at least a factor of 4, in turn showing a quantum
advantage for such thermalization problems at a
factor of 4 smaller qubit number for open quantum systems as opposed
to closed ones. Thus, our results encourage open quantum system
exploration in noisy intermediate-scale quantum technologies.

\end{abstract}

\maketitle

%\tableofcontents

% =============================================================================
\section{Introduction                                                          \label{sec:tqi:intro}}
% =============================================================================
%

% Initial paragraph, thermalization hypothesis, outline introduction

In statistical mechanics, it is assumed that a
many-body quantum system coupled to a thermal bath will thermalize
independently of the initial state. Nevertheless, the conditions for
thermalization are subtle and need to be questioned for each specific
setting \cite{Eisert2015,Gogolin2016}. For example, the presence of gaps in the
density of states of the bath \cite{Xiong2015}, the existence of symmetries
\cite{Stobinska2014}, or the emergence of dynamical phase transitions in the
thermodynamic limit may affect the thermalization process. Understanding
dissipation and thermalization is crucial for quantum technologies, and
includes aspects such as the engineering of steady state preparation or the
control of quantum decoherence.

The simplest approach to describe the evolution of a quantum
system coupled to a thermal bath is the Lindblad master equation
\cite{Kossakowski1972,Lindblad1976,Gorini1976,Rivas2012,BreuerPetruccione},
which preserves mathematical properties of the open system density matrix
such as positivity and norm.
With this framework, the description of a single-body open system is
straightforward, but many-body systems impose challenges that push
numerical methods to their performance limit. Indeed, to correctly
describe thermalization, the master equation requires an exact
diagonalization of the system, such that all transitions between
its eigenstates are considered. Hence, the Lindblad equation contains
as many terms (or decaying channels) as transitions between the system
energy levels.
A further simplification is to consider single-channel Lindblad
equations, which depend on Lindblad operators that act locally
or quasi-locally on each qudit forming the open system. This approach
is accurate for situations to describe steady states or when the bath
produces a single relevant transition, for
instance in weakly interacting qubits. It has been used to analyze transport
in spin problems, where a one-dimensional many-body open system couples to
baths via both the first and last site \cite{Prosen2008,Prosen2009,Prosen2011,Palmero2019} or
to study decoherence in strongly correlated systems
\cite{Cai2013,Daley2014,Rota2017,Bernier2018}. However,
as shown in \cite{Hofer2017,Smith2017,Gonzalez2017,Raja2018} for single open
qubits or qudits, the single-channel approach may not be sufficiently accurate to
describe thermalization or even quantum transport \cite{Mascarenhas2018}.
A qualitative reason is that to describe the relaxation of a many-body,
or even a multi-level, open system to its thermal state requires the bath
redistribute the occupation probabilities of all its energy levels
according to a Gibbs distribution. Naturally, this
process can only be accounted for correctly when the Lindblad equation
produces transitions among all energy levels of the system.

In this work, we consider a harmonic bath weakly coupled to our system. To
justify our use of the Lindblad equation, a Born-Markov approximation is
considered, in which we take the relaxation
timescale of the bath to be very short in comparison to the relaxation
timescale of the system. This approximation implies that the bath state
only suffers small fluctuations around its thermal initial state, such
that only short-time system-bath correlations are considered. A further
step to derive the Lindblad equation is to consider the secular approximation,
which neglects processes involving large differences between two energy
transitions, and which is based on a similar argument as used for the
rotating wave approximation.

We further consider that our open system is described by the
well-known quantum Ising model~\cite{Ising1925,SachdevQPT}.
Its equilibrium and dynamical properties have been widely analyzed,
including generalizations such as long-range
interactions \cite{Dutta2001,Koffel2012,Vodola2015,JaschkeLRQIC}.
The open quantum system dynamics of 1D spin chains have been studied
in an end-cap scenario with the idea of a wire attached
to either end, or else a subsystem of a much larger 1D spin chain.
This approach amounts to independent baths coupled to both ends of
the chain with Lindblad operators constructed in terms of Hermitian
Majorana operators  \cite{Prosen2008}, or defined in terms of the
first two and last two spins as in \cite{Prosen2009}, a proposal
recently extended in \cite{Palmero2019}. The basic aim of these
proposals is to drive the system to its thermal state, by imposing
a local version of the detailed balance condition. Alternatively,
\cite{Vogl2012} analyzes the same configuration by constructing the
Lindblad operators in terms of the system eigenstates, which are
computed by assuming periodic boundary conditions.
However, a general description of not only the steady state but also
the relaxation process of an Ising chain coupled to a thermal bath in
different configurations and without periodic conditions
has not been done up to now to our knowledge.
Such analysis is particularly timely in quantum simulation experiments
where, among other spin systems, e.g., the $XYZ$ model
\cite{carr2013j,carr2015e}, as well as Bose- and Fermi-Hubbard models,
the Ising model is one of the main models of choice~\cite{lewensteinM2007}.
Trapped ions and Rydberg chains either in optical lattice or optical
tweezer arrays can produce Ising chains~\cite{Bohnet2016,Labuhn2016},
and generalized Ising models with different couplings such as the
chimera map in D-Wave form the basis of practical quantum machines.
Thus, understanding the thermalization and decoherence properties of
the minimal 1D nearest-neighbor Ising chain is an essential
first step. We build this description and extend previous studies in two ways.

% Problems addressed and novelty
%
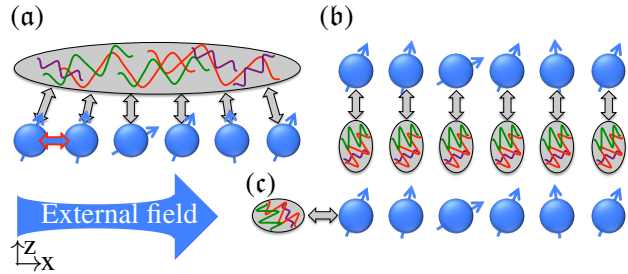
\begin{figure}[t]
  \begin{center}
    \vspace{0.5cm}
    \begin{overpic}[width=0.5 \columnwidth,unit=1mm]{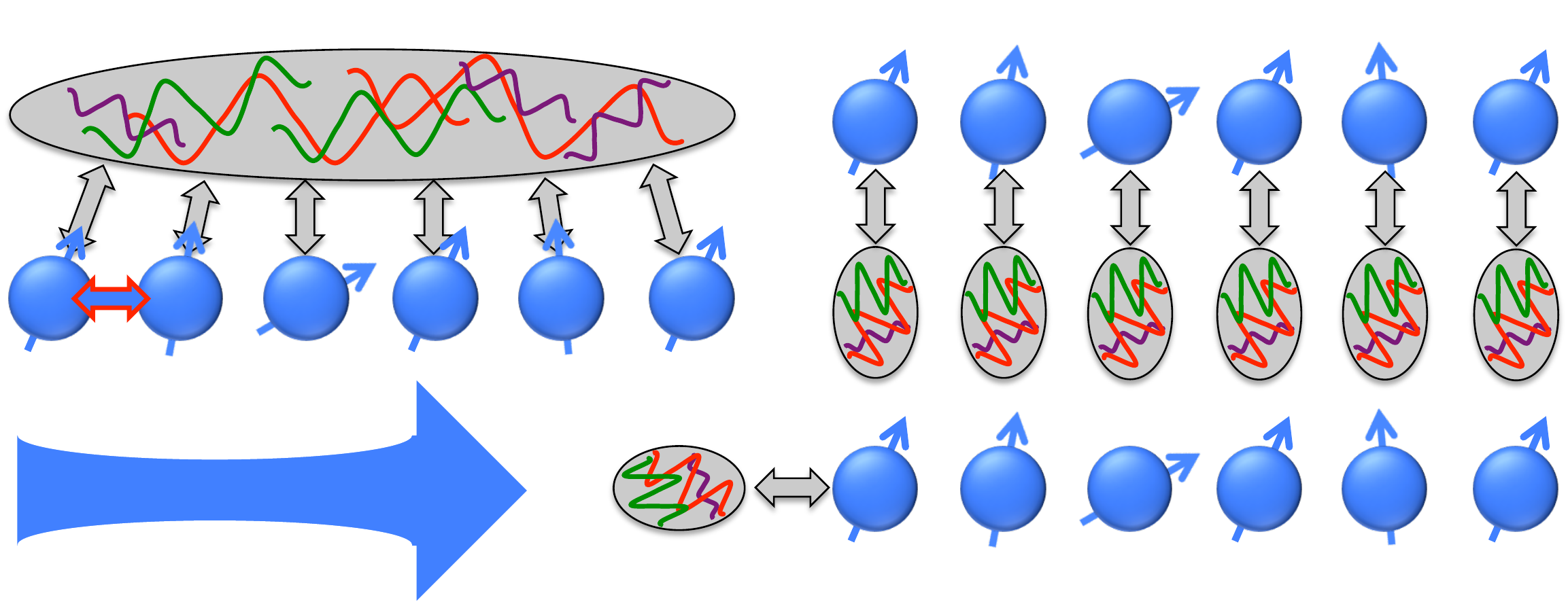}
      \put( 0,38){\srcasea}
      \put( 50,38){\srcaseb}
      \put( 38,11){\srcasec}
      \put( 5,6){\color{white}External field}
      \put( 0,0){$\uparrow$}
      \put( 0.6,-1.75){$\rightarrow$}
      \put( 4.9,-1.5){x}
      \put( 2.1,1){z}
    \end{overpic}
    \caption[Spins in the quantum Ising model coupled to a bath]
      {\emph{Quantum Ising chain coupled to three different baths: common, independent, and end-cap.}
      The thermalization of the quantum Ising model when coupled to a thermal
      bath raises many intriguing questions: Will the system thermalize? If so, at what rate?
      Are there any suitable approximations to model the process in the many-body limit? The system Hamiltonian consists of an
      external field in the $x$-direction indicated by the big blue arrow
      trying to align the spin $1/2$ particles (blue spheres with arrow) in the
      $x$-direction to reach the ground state. The nearest-neighbor interaction
      in the $z$-direction, highlighted only for the first two sites in \srcasea{} competes with the external field in the ground state trying to align neighboring spins.
      The baths correspond to a three-dimensional electromagnetic
      field represented by grey ovals with a dense and wide mode spectrum.
      The grey arrows indicate the interaction with each of the system spins.
      We consider: \srcasea{} a \emph{common} bath coupled to all sites,
      \srcaseb{} \emph{independent} baths for each site, and \srcasec{}
      an end-cap scenario in which only the first site is coupled to the bath. Multi-channel Lindblad
      operators can simulate all three scenarios; neighborhood
      single-channel Lindblad operators are enabled for scenarios \srcaseb{}
      and \srcasec{}.
                                                                                \label{fig:tqi:schematic}}
  \end{center}
\end{figure}

First, we analyze three experimentally feasible, distinct scenarios of
coupling a bath to an Ising chain as presented in Fig.~\ref{fig:tqi:schematic}:
These are \srcasea{}~a \textit{common} bath coupled to all sites,
thus producing atom-atom interactions,
\srcaseb{}~\textit{independent} baths coupled to each site, and \srcasec{}~a
single \textit{end-cap} bath coupled only to the first site in the Ising
chain. We determine the resulting timescale for thermalization and discuss how
crucial it is to consider the effects of the $\mathbb{Z}_{2}$ symmetry of the quantum
Ising model. For our choice of the interaction Hamiltonian, we find that the
Lindblad description preserves the symmetry sectors and, thus, the system thermalizes
in each symmetry sector, but not across sectors. This statement excludes results from
Secs.~\ref{sec:resthermsingleferro} and \ref{beyond}.

Second, we compare a multi-channel Lindblad equation to single-channel
Lindblad approaches. The multi-channel case makes no further
approximations to the microscopically derived Lindblad equation and,
therefore, it correctly describes the system thermalization within such
a framework. However, tensor network methods such as matrix product
density operators (MPDO) cannot be applied. The multi-channel
approach reduces by at least a factor of four the number of qubits
that can be treated with respect to the closed quantum system case. Indeed,
while exact diagonalization of the Hamiltonian of closed systems is
limited to 14 qubits on typical personal computers or laptops, the time evolution
of an initial state can be extended to larger systems. Without
using parallelization, Krylov and Trotter methods allow for the evolution
of at least 24-25 qubits, while parallelization techniques allow one
to treat closed systems as quantum circuits of 45 to 64
qubits~\cite{Haner2017,Pednault2017,Chen2018}.
Hence, the single-channel local approach is commonly used for many-body
calculations, since it allows one to employ the same techniques as for
the closed system case when evolving a density matrix instead of a pure state.
Moreover, MPDO is a highly efficient method to reach much larger systems.
To be able to describe thermalization while at the same time keeping the
Lindblad equation as local as possible, we propose quasi-local or three-site
neighborhood Lindblad operators, where two adjacent spins control the flip
of the central one. This operation is known from closed systems as
quantum gates, in our case a three qubit
extension of the standard CNOT (controlled-not) gate, where an operation
on one qubit is based on the state of another qubit.  However, we will
show that the neighborhood approach only produces accurate results in
the extreme paramagnetic or ferromagnetic limits. For the Ising chain
anywhere in the proximity of the quantum phase transition, the vital
factor of four overhead required for the multi-channel calculation is
necessary. In other words, we show that a quantum advantage occurs for
at least a factor of four less qubits for open quantum
systems as compared to closed ones. For instance, if we supposed
quantum advantage for 50 qubits in a closed system, for an open quantum
system we only require around 13 qubits. Our results
put the emphasis on open quantum system quantum simulators for the
nearest term noisy intermediate-scale quantum (NISQ) technology.

% Outline
%
The outline of our work is as follows. We introduce the closed and open
quantum systems in Sec.~\ref{model}. We continue in
Sec.~\ref{sec:tqi:lopchoice} with a detailed look at our evolution
equations.  This section includes multi-channel and single-channel Lindblad
equations and their properties, such as the eigenvalue decomposition and
symmetries.
Then, we evaluate the thermalization timescales for different scenarios in
Sec.~\ref{sec:tqi:results}. Finally, we discuss possible experimental
realizations in Sec.~\ref{experiment} and conclude our results in
Sec.~\ref{sec:tqi:concl}. Some topics we have included as appendices
for the interested reader who wishes to consider in detail additional
aspects supporting the main study: the odd symmetry sector in
App.~\ref{app:tqi:odd}; details on the coupling strengths in the
multi-channel Lindblad equation as well as in the derivation of the single-channel
operators in App.~\ref{sec:tqi:full}; and the choice of a different
operator in the interaction Hamiltonian in App.~\ref{app:tqi:sz}.

% =============================================================================
\section{Ising Spin Chains and their Bath Couplings}                           \label{model}
% =============================================================================

In this section, we remind the reader of the system Hamiltonian, namely the
transverse quantum Ising model. We then introduce the different system-bath configurations
as well as the models of Lindblad master equation that we will consider.
We discuss the common, independent, and end-cap bath scenarios \srcasea{},
\srcaseb{}, and \srcasec{} from Fig.~\ref{fig:tqi:schematic}.

% =============================================================================
\subsection{Transverse quantum Ising model                                     \label{sec:tqi:closedsystem}}
% =============================================================================
%
We focus on the transverse quantum Ising model in one dimension with a Hamiltonian given as
\begin{eqnarray}                                                                \label{eq:tqi:HQI}
  H_{\mathrm{QI}}
  &=& - \sum_{k=1}^{N-1} \cos(\phi) \sigma_{k}^{z} \sigma_{k + 1}^{z}
      - \sum_{k=1}^{N} \sin(\phi) \sigma_{k}^{x} \, ,
\end{eqnarray}
where the formulation of the coupling in terms of sine and cosine simplifies
addressing the ferromagnetic ($\phi = 0$) and paramagnetic ($\phi = \pi / 2$)
limit. The system size or the number of qubits in the
quantum Ising chain is $N$. The Pauli spin operators $\sigma_{k}^{x}$ and
$\sigma_{k}^{z}$ are acting on site $k$, $\cos(\phi) \equiv J
/ \sqrt{J^2 + g^2} > 0$ is the unitless interaction energy, and $\sin(\phi)
\equiv g / \sqrt{J^2 + g^2}$ is the unitless
coupling to the external field. The critical point of the quantum Ising
model is at $\phi_{c} = \pi / 4$ for the thermodynamic limit $N \to \infty$,
i.e., $\sin(\phi_{c}) = \cos(\phi_c) = 1 / \sqrt{2}$. We treat systems below ten qubits
and the position of the minimum of the gap $\phi_{c}$, i.e., an indicator of
the quantum critical points, shifts to values $\phi_{c} < \pi / 4$ due to
finite-size effects. Note that in this formulation the Hamiltonian
$H_{\mathrm{QI}}$ is expressed in terms of the $\sqrt{J^2 + g^2}$, and
all quantities are therefore unitless.

The quantum Ising model has a $\mathbb{Z}_{2}$ symmetry. This symmetry is
very useful when treating the closed quantum system, since it leads to a
block-diagonal structure of the Ising Hamiltonian in  two blocks representing
respectively the even and odd sectors. Furthermore, most of our results are
based on choosing an interaction operator $S_{k} = \sigma_{k}^{x}$ that
preserves the symmetry, and thus can treat the
Lindblad master equation in a particular symmetry sector.
In contrast, the choices $S_k=\sigma_{k}^{z}+\sigma_k^x$ discussed in
Sec.~\ref{beyond} and $S_k=\sigma_k^z$ discussed  in App.~\ref{app:tqi:sz} do
not conserve the symmetry. Reference~\cite{Buca2012} presents
a more detailed discussion of symmetries in the Lindblad master equation.

In the following, we explore the quantum Ising system coupled to a dissipative bath
for the common, independent, and end-cap configurations.

% ..............................................................................
\subsection{Coupling to the electromagnetic field in three different configurations}
% ..............................................................................

First, we consider the case
in which all spins are coupled to a common bath of harmonic oscillators, as
described in Fig.~\ref{fig:tqi:schematic}\srcasea{}, such that the total
Hamiltonian of the system (S) and the bath (R) can be written as
\begin{eqnarray}                                                                \label{eq:tqi:HSR}
  H^{\msrcasea}_{\mathrm{\mathrm{S+R}}}
  &=& H_{\mathrm{QI}}
  + \sum_{k=1}^{N} \sum_{\bf q}  S_{k} (g_{k{\bf q}} b_{{\bf q}} + g_{k{\bf q}}^{\ast} b_{{\bf q}}^{\dagger})
  + \sum_{\bf q} \omega_{q} n_{{\bf q}} \, ,
\end{eqnarray}
where $b_{\bf q}^{\dagger}$ ($b_{\bf q}$) is the creation (annihilation) operator of the
bath with a frequency $\omega_{q}$, and momentum ${\bf q}$. Also,
$n_{{\bf q}} = b_{\bf q}^{\dagger} b_{\bf q}$ is the number operator.

% ..............................................................................
%\subsection{Local bath}
% ..............................................................................

As a second configuration, we consider that each spin is coupled to its
independent bath, as described in
Fig.~\ref{fig:tqi:schematic}\srcaseb{}. The total Hamiltonian is
\begin{eqnarray}                                                                \label{eq:tqi:HSRlocal}
  H^{\msrcaseb}_{\mathrm{\mathrm{S+R}}}
  &=& H_{\mathrm{QI}}
  + \sum_{k=1}^{N} \sum_{\bf q} g_{1{\bf q}}  S_{k} (b_{kq} + b_{kq}^{\dagger})
  + \sum_{k=1}^{N} \sum_{\bf q} \omega_{q} n_{k{\bf q}} \, , \nonumber \\
\end{eqnarray}
where now there is a set of harmonic oscillators $b_{kq}^{\dagger}$
corresponding to the bath attached to each spin $k$.

% ..............................................................................
%\subsection{End-cap bath}
% ..............................................................................

Finally, the configuration in Fig.~\ref{fig:tqi:schematic}\srcasec{} corresponds to
the case described in Eq.~\eqref{eq:tqi:HSR} with $g_{k{\bf q}}=0$ for every
$k \neq 1$. The interaction Hamiltonian is then reduced to
\begin{eqnarray}                                                                \label{eq:tqi:HSRfirst}
  H^{\msrcasec}_{\mathrm{S+R}}
  &=& H_{\mathrm{QI}}
  + \sum_{\bf q} g_{1q} S_{1} ( b_{1{\bf q}} + b_{1{\bf q}}^{\dagger})
  + \sum_{\bf q} \omega_{q} n_{1{\bf q}} \, .
\end{eqnarray}
In all three configurations, the interaction between the system and the
bath is modulated by the coupling strength $g_{k{\bf q}}$. For the
three-dimensional electromagnetic field considered here, such coupling
strengths correspond to a dipolar  coupling \cite{Lehmberg1970},
\begin{eqnarray}                                                                \label{eq:EMfield}
  g_{k{\bf q}} &\sim&  \sqrt{\hbar\omega_q / (\nu \epsilon_{0} )} \;
  (\vec{r}_{k} \cdot \vec{d}_{\mathrm{dip}}) / r_{k}
  \exp(- \mathrm{i} \vec{r}_{k} \cdot \vec{q}) \,
\end{eqnarray}
where $\nu$ is the quantization volume. We consider the Planck constant
$\hbar$ and the electric permittivity $\epsilon_{0}$ as unit constants.
Furthermore, the coupling $g_{k{\bf q}}$ contains the scalar product
between the position of each spin $\vec{r}_{k}$ and the dipole vector
$\vec{d}_{\mathrm{dip}}$, and a different phase factor $\exp(- \mathrm{i}
\vec{r}_{k} \vec{q})$ for each position and each momentum. Here,
we assume that all spins have equal dipole moments and that these are
orthogonal to the atomic positions. Atoms are regularly arranged in a
chain along the $x$ axis, such that
${\bf r}_{k}=(k-1)a_{\mathrm{lat}}{\bf \hat{x}}$, with $a_{\mathrm{lat}}$ the
between adjacent spins in the chain.
We note that the same physics
described with Eq.~\eqref{eq:tqi:HSRlocal} can be obtained with the Hamiltonian
in Eq.~\eqref{eq:tqi:HSR}, i.e., with a single bath, under particular
conditions. This situation is achieved under the condition
$q^\mathrm{min}_0 a_{\mathrm{lat}}\ll 1$, where
$q_{0}^{\mathrm{min}}$ is the resonant wave vector of the field with
smallest modulus. Resonant wave vectors correspond
to bath modes that interact resonantly with the system frequencies
$\omega$, such that $\omega_{q_0}=\omega$. This  condition ensures the
absence of bath-mediated dipole-dipole interactions between the spins,
which therefore evolve as through each of them was coupled to their
independent bath.

The interaction Hamiltonian also depends on the system coupling
operators $S_{k}$, which in most of our examples we  choose as
$S_{k} = \sigma_{k}^{x}$. We point out that the paramagnetic limit
$\phi = \pi / 2$ corresponds to a non-interacting
model, a case that has been extensibly discussed in \cite{Ostilli2017}.

% =============================================================================
\section{Different models of Lindblad equations}                               \label{sec:tqi:lopchoice}
% =============================================================================

Let us briefly point out in this section the difference between the global
multi-channel approach and neighborhood single-channel approach. We use the
same Lindblad master equation for both approaches, which we define as
\begin{eqnarray}                                                               \label{eq:tqi:lindblad}
  \dot{\rho} &=& - \frac{\mathrm{i}}{\hbar} \left[ H, \rho \right]
  + \sum_{abcd} \mathcal{C}_{abcd} \left( L_{ab} \rho L_{cd}^{\dagger}
  - \frac{1}{2} \left \{ L_{cd}^{\dagger} L_{ab} , \rho \right \} \right)
  \qquad \Longleftrightarrow \qquad
  \frac{\partial}{\partial t} \Lket{\rho} = \mathcal{L} \Lket{\rho} \, ,
\end{eqnarray}
where the formulation with the density matrix $\rho$ (left hand side) or with the
super-ket $\Lket{\rho}$ in the Liouville space (right hand side) are equivalent.
Equation~\eqref{eq:tqi:lindblad} evolves the density matrix $\rho$ of a shape
$D \times D$, where $D$ is the dimension of the Hilbert space of the quantum
Ising chain. In contrast, the vector representation $\Lket{\rho}$ of the density matrix
has $D^2$ entries while the matrix $\mathcal{L}$ acting on $\Lket{\rho}$ from the left-hand
side is of dimension $D^2 \times D^2$ due to the outer product or Kronecker
product, e.g., $H \otimes I$ and $I$ is an identity matrix of size $D \times D$. The
Liouville operator is defined as
\begin{eqnarray}                                                                \label{eq:tqi:liouvillian}
  \mathcal{L} &=& - \frac{\mathrm{i}}{\hbar}
  \left( H \otimes \1 - \1 \otimes H^{T} \right)
  + \sum_{abcd} \frac{\mathcal{C}_{abcd}}{2} \left( 2 L_{ab} \otimes L_{cd}^{\ast}
  - L_{cd}^{\dagger} L_{ab} \otimes \1
  - \1 \otimes L_{ab}^{T} L_{cd}^{\ast}
   \right) \, ,
\end{eqnarray}
In the Lindblad equation, the Hamiltonian
$H$ corresponds to the Hamiltonian of the system and possible corrections from the
system-bath interaction. $L_{ab}$ are the Lindblad operators encoding
the dissipative part of the dynamics via a transition from state $\ket{b}$
to $\ket{a}$ with a coupling $\mathcal{C}_{abcd}$. Thus, the sum over $a$,
$b$, $c$, and $d$ iterates over a set of states.
The superscripts $T$ and $\ast$ are the transpose and complex conjugate
of an operator, respectively.
Throughout the work, we choose our units such that $d_{\mathrm{dip}} = 1$,
the lattice spacing $a_{\mathrm{lat}} = 1$ between sites, the speed
of light $c = 1$, the Boltzmann constant
$k_\mathrm{B} = 1$, and $\hbar=1$. Thus, the temperature $T$ is
implicitly in units of $[J^2 + g^2]^{1/2}$ and times are given implicitly in
units of $[J^2 + g^2]^{-1/2}$.
The choice of the Lindblad operators and decay rates depends on the particular
case considered. We shall consider two different cases, a multi-channel
Lindblad equation as derived from Markov and secular approximations
\cite{BreuerPetruccione}, and a single-channel equation with neighborhood
single-channel Lindblad operators. These options are detailed in the following.

% ..............................................................................
\subsection{Multi-channel Lindblad equation                                  \label{sec:tqi:choicemulti}}
% ..............................................................................

In this first approach, we consider the master equation obtained from first
principles according to the standard derivation, see
\cite{BreuerPetruccione,Rivas2012,schaller2014} for details. This path considers
the well-known Born-Markov and secular approximations. A crucial point of this
derivation is to diagonalize the system Hamiltonian, such that
\begin{eqnarray}
  H_{S} &=& \sum_{a} E_{a} \ket{a} \bra{a} \, .
\end{eqnarray}
The sum over $a$ contains $d_{\mathrm{loc}}^{N}$ eigenstates without
$\mathbb{Z}_{2}$ symmetry, and $d_{\mathrm{loc}}^{N-1}$ eigenstates
within a symmetry sector; $d_{\mathrm{loc}}$ is the local dimension.
This definition allows one to re-express the
system coupling operators $S_k$ in terms of
such eigenstates,
\begin{eqnarray}
  S_k=\sum_{ab}\langle a|S_k|b\rangle L_{ab}.
\end{eqnarray}
In this representation, the resulting master equation is written in terms of
Lindblad operators that represent transitions between different eigenstates,
$\ket{b}$ to $\ket{a}$,
\begin{eqnarray}
  L_{ab} &=& \ket{a} \bra{b} \,,
\end{eqnarray}
and therefore, the number of Lindblad operators is $d_{\mathrm{loc}}^{2N}$
where $d_{\mathrm{loc}} = 2$ in our case \footnote{This
upper bound could, in theory, be reduced to $d_{\mathrm{loc}}^{2N} - 1$,
but is not of interest for numerical implementation.}.
Notably, the energy eigenstates can be entangled in the computational basis, i.e.,
the Fock basis corresponding to the qubits; thus, the state of any spin $k'$ can
have different values in the eigenstates $\ket{a}$ and
$\ket{b}$ and change during the action of $L_{ab}$, although the operator
in the interaction Hamiltonian acts on site $k$ via $S_{k}, k \neq k'$.
This property contributes to their being named global Lindblad operators.

The decay rates $\mathcal{F}_{abcd}$ in Eq.~\eqref{eq:tqi:lindblad} can be written as
\begin{eqnarray}                                                               \label{eq:tqi:ratederiv1}
  \mathcal{C}_{abcd} = \sum_{j=1}^{N} \sum_{k=1}^{N} \mathcal{C}_{abcd}^{[jk]} \,
  \, , \qquad
  \mathcal{C}_{abcd}^{[jk]} &=& \delta_{E_{ba} - E_{dc}} \gamma^{[jk]}(E_{ba},E_{dc}).
\end{eqnarray}
where $E_{ba}=E_b - E_a$ and $E_{dc}=E_d - E_c$ correspond to the two system
transitions that are present in each term of the equation. However, the
secular approximation considered forces such transitions to correspond
to the same  frequency, as enforced by the Kronecker delta in
Eq.~\eqref{eq:tqi:ratederiv1}. Moreover, the term $\gamma^{[jk]}(E_{ba},E_{dc})$
depends on the matrix elements of the system coupling operator in the two
system transitions,
\begin{eqnarray}                                                               \label{eq:tqi:ratederiv2}
\gamma^{[jk]}(E_{ba},E_{dc})
  &=& \bra{a} S_{j} \ket{b} \bra{d} S_{k} \ket{c} \gamma(E_{ba}, {\bf r}_{jk}) \, .
\end{eqnarray}
where ${\bf r}_{jk}$ represents the distance between the two spins $i$ and
$j$. This dependency describes the fact that the bath dissipation connects
different spins with each others, which gives rise to collective or
cooperative effects in the relaxation process. Furthermore, we note
that the function $\gamma(E_{ba}, {\bf r}_{jk})$ is specific to each type
of bath and carries the dependency of the temperature. Here, the bath is
a three dimensional electromagnetic field, to which the Ising spins are
coupled via the standard dipolar interaction as described by the
Hamiltonians~\eqref{eq:tqi:HSR}, \eqref{eq:tqi:HSRlocal}, and
\eqref{eq:tqi:HSRfirst} in the three configurations discussed, with
the couplings strengths defined in Eq.~\eqref{eq:EMfield}. Because the bath is in the thermal
equilibrium, the decay rates fulfill the detailed balance condition, and
therefore the steady state of the equation is the thermal state.

Further details of this discussion can be found in App.~\ref{sec:tqi:full}.

% ..............................................................................
\subsection{Single-channel Lindblad equation}
% ..............................................................................

We propose single-channel Lindblad operators that allow us to express the
action of the bath in terms of a number of decaying channels
that scales linearly with the number of qubits only. This advantage, together with
the fact that they act only either on a qubit or its neighborhood, make them
computationally more feasible.
Hence, single-channel Lindblad operators are very useful to describe dissipation
in many-body systems without considering the complete energy spectrum,
while still transitioning between energy eigenstates in the corresponding
limits.

Our design of these operators is inspired by the spectral decomposition described in
the previous section. However, the idea is to concentrate on the action of
the operator $S_{k}$ in $\ket{a} \bra{a} S_{k} \ket{b}
\bra{b}$ on a single site $k$. To this aim, we use the Schmidt
decomposition \cite{NielsenChuang} and rewrite the
eigenstates $\ket{a}$ and $\ket{b}$ in terms of a bipartition between the
site $k$ and all others sites $\bar{k}$, i.e., all $k'$ with $k' \neq k$,
\begin{eqnarray}
  \ket{a} = \sum_{i} \lambda_{ika} \ket{a_{k}^{i}} \ket{a_{\bar{k}}^{i}} \, , \qquad
  \ket{b} = \sum_{j} \lambda_{jkb}' \ket{b_{k}^{j}} \ket{b_{\bar{k}}^{j}} \, .
\end{eqnarray}
The Schmidt decomposition characterizes the entanglement between two
subsystems of a pure quantum state and generates two orthonormal sets of
basis states, $\{\ket{a_{k}^{i}}\}$ and $\{\ket{a_{\bar{k}}^{i}}\}$ for each
bipartition. The number of components in the Schmidt decomposition
depends on the system size and
amount of entanglement the state and is therefore not specified.
The weight of each basis state within the
decomposition is set via the singular values $\lambda_{ika}$, which are
larger than or equal to zero by definition. With such a decomposition,
we can write
\begin{eqnarray}                                                                \label{eq:tqi:schmidtfull}
  \ket{a} \bra{a} &S_{k} &\ket{b} \bra{b}
  = \sum_{ij'}^{} \, ^{ij'}\mathcal{F}_{ab}^{[k]} \ket{a_{k}^{i}} \bra{b_{k}^{j'}}
      \otimes \ket{a_{\bar{k}}^{i}} \bra{b_{\bar{k}}^{j'}} \, , \nonumber \\
  ^{ij'}\mathcal{F}_{ab}^{[k]} &=& \lambda_{ika} \lambda_{j'kb}' \bigg( \sum_{i'j}
  \lambda_{i'ka} \lambda_{jkb}' \bra{a_{k}^{i'}} S_{k} \ket{b_{k}^{j}} \cdot
  \braket{a_{\bar{k}}^{i'}}{b_{\bar{k}}^{j}} \bigg) \, .
\end{eqnarray}
The coefficient $^{ij'}\mathcal{F}_{ab}^{[k]}$ depends on the eigenstates
$a$ and $b$ of the system Hamiltonian $H_{\mathrm{QI}}$.
Furthermore, the orthonormal basis states
of the Schmidt decomposition are denoted by superscripts $ij'$ in front
of the symbol.

Building the Lindblad local operators is particularly feasible in the
ferromagnetic and paramagnetic limits, since in these limits the
eigenstates are not entangled and the Schmidt decomposition contains a
single term. In the paramagnetic limit, to observe thermalization we
have to choose a coupling operator that is not diagonal to the paramagnetic
Hamiltonian, for instance, $S_{k} = \sigma_{k}^{z}$. The corresponding
Lindblad operator corresponds to a local spin flip which changes by a
quanta the energy of the system. A complete numerical analysis of the
dynamics for such a case is presented in App.~\ref{app:tqi:sz}. In the
ferromagnetic limit, the Lindblad operators have to act on three qubits
in order to describe a transition that produces an energy change;
we are using the symmetry-broken eigenstates to avoid entanglement
in the ferromagnetic limit and, thus, do not use the $\mathbb{Z}_{2}$
symmetry.
More details on the structure and nature of single-channel Lindblad
operators is given in App.~\ref{sec:tqi:limits}.

% -----------------------------------------------------------------------------
\subsection{Eigenstate decomposition of the Lindblad equation                                \label{sec:tqi:prop}}
% -----------------------------------------------------------------------------

It can be analytically shown that a steady state or fixed point of the
multi-channel Lindblad equation is the thermal state \cite{BreuerPetruccione}
\begin{eqnarray}                                                                \label{eq:tqi:thermal_global}
  \rho_{\mathrm{th}} &=& \frac{\exp\left(- \beta H_{S} \right)}{ Z} \, ,
\end{eqnarray}
with $\beta =1 / (k_{B} T)$ and the partition function $Z = \mathrm{Tr} \left[ \exp\left(
- \beta H_{S} \right) \right]$, and $H_S$ is the Ising model in our case.
However, even with a multi-channel Lindblad equation, the system may not relax
to such a thermal state. To analyze this thermalization as well as the decay rates
of the system, we take the Lindblad equation in its
vector form Eq.~\eqref{eq:tqi:lindblad}, and diagonalize
the Liouville operator. Since it is not Hermitian, its
diagonalization gives rise to left and right eigenvectors. The right
eigenvectors, denoted as $\Lket{v_{j}}$, fulfil the condition
$\mathcal{L} \Lket{v_{j}} = \Lambda_{j} \Lket{v_{j}}$, where $\Lambda_{j}$
are the corresponding eigenvalues of the Liouville operator $\mathcal{L}$.
Thus, the eigenvectors $\Lket{v_j}$ form a complete basis and we can express
any initial density matrix as a linear combination of the basis vectors with
some complex weight $c_{j}$, i.e., $\Lket{\rho(0)} = \sum_{j}
c_{j} \Lket{v_{j}}$. The time-evolved state can then be written
as \cite{Albert2014,manzano2018}
\begin{eqnarray}                                                                \label{eq:tqi:lioutimeevo}
  \Lket{\rho(t)} &=& \sum_j c_j \mathrm{e}^{\Lambda_j t} \Lket{v_{j}} \, ,
\end{eqnarray}
where $c_j=\langle\langle {v}_j|\rho(0)\rangle\rangle$. On the one hand, the
terms corresponding to eigenvalues $\Lambda_{j}$ with zero real part do
not decay over time and therefore correspond to steady states. Moreover,
if there is a single eigenstate with zero real part eigenvalue, this
corresponds to the thermal state. On the other hand, the contribution of
eigenvectors with a negative real part will vanish within a timescale
given by $|\Re(\Lambda_j)|$, where $\Re$ refers to the real part.

If we sort the eigenvalues as $|\Re(\Lambda_{j})| \le |\Re(\Lambda_{j+1})|$,
and find only the first one ($\Lambda_{0}$) has a zero real part, the
second one $|\Re(\Lambda_1)|$ will define a gap that describes the
slowest decay timescale of the system. Further details of the Liouville
spectrum have been discussed in
literature \cite{Sarandy2005,Buca2012,Horstmann2013,Albert2014,manzano2018}.

% ..............................................................................
\subsection{Effects of symmetries}                                              \label{symmetries}
% ..............................................................................

As mentioned before, the Ising model has a
$\mathbb{Z}_{2}$ symmetry that is preserved by the dissipation and divides the
Hilbert space into even and odd symmetry sectors. We choose our interaction
operator accordingly to ensure the conservation of the symmetry.
For simplicity we will perform our
analysis within the even sector, which means that instead
of analyzing thermalization to a state
Eq.~\eqref{eq:tqi:thermal_global}, we analyze thermalization to
\begin{eqnarray}                                                                \label{eq:tqi:thermal_even}
  \rho_{\mathrm{th}}^{\mathrm{e}}
  &=& \frac{\exp\left(- \beta H_{S}^{\mathrm{e}} \right)}{Z} \, ,
\end{eqnarray}
with $H^{\mathrm{e}}_{S}$ the Ising Hamiltonian in the even sector.
We briefly discuss the dynamics within the odd symmetry section
in App.~\ref{app:tqi:sz}. The use of the symmetry has two advantages.

Firstly, if we do not use the symmetry, we have at
minimum two steady states, i.e., one for each
symmetry sector. This would give rise to a degeneracy in the steady state
eigenstates of the Liouville operator, and therefore to a more complex
analysis of the properties of the steady state.
Secondly, the symmetry improves the computational scaling. For
our spin system, the size of the Liouville operator decreases from $2^{2N} \times
2^{2N}$ to $2^{2(N - 1)} \times
2^{2(N - 1)}$ which simplifies computational
operations such as the eigenvalue decomposition. Therefore, we also have
only two sets of $2^{2(N - 1)}$ eigenvalues and eigenstates
instead of $2^{2N}$.

% =============================================================================
\section{Thermalization timescales                                             \label{sec:tqi:results}}
% =============================================================================

In the following, we analyze the time evolution of the system and the spectrum
of the Liouville operators $\mathcal{L}$. We execute this analysis for the
different scenarios of a system-bath coupling sketched in
Fig.~\ref{fig:tqi:schematic} and represented by the Hamiltonians in Eqs.~\eqref{eq:tqi:HSR} to
\eqref{eq:tqi:HSRfirst}. We find the timescale for the thermalization
of the system for the multi-channel Lindblad operators.

% -----------------------------------------------------------------------------
\subsection{Numerical Methods}
% -----------------------------------------------------------------------------

Our results are based on the exact diagonalization
obtained by using the OSMPS package \cite{JaschkeED,OpenMPS}.
Without considering symmetries, the number of transitions and decay
channels in the Lindblad equation, as well as the
dimension of the density matrix, grows in the number of spins as $2^{N}$.
Besides considering the time evolution, we analyze the spectrum
of the Liouville propagator and study steady
states and timescales characterized by the gap in the Liouville
propagator.
Such an analysis requires diagonalizing a square
matrix of dimension $2^{2N}$, which is even more limiting. Hence, calculations
beyond seven qubits are infeasible without parallelization, and
each additional qubit in the open systems increases the computational
resources used by a factor of $64$ due to the cubically scaling eigenvalue
decomposition. As discussed in the previous section, this scaling is
partially improved by the fact that we will be working in the even
sector of the $\mathbb{Z}_{2}$ symmetry.

\begin{figure}[t]
  % /data/FinalSpinsInEM3D/01_Thermal_Scales
  \begin{center}
    \vspace{0.5cm}
    \begin{minipage}{0.47\linewidth}
      \begin{overpic}[width=0.9 \columnwidth,unit=1mm]{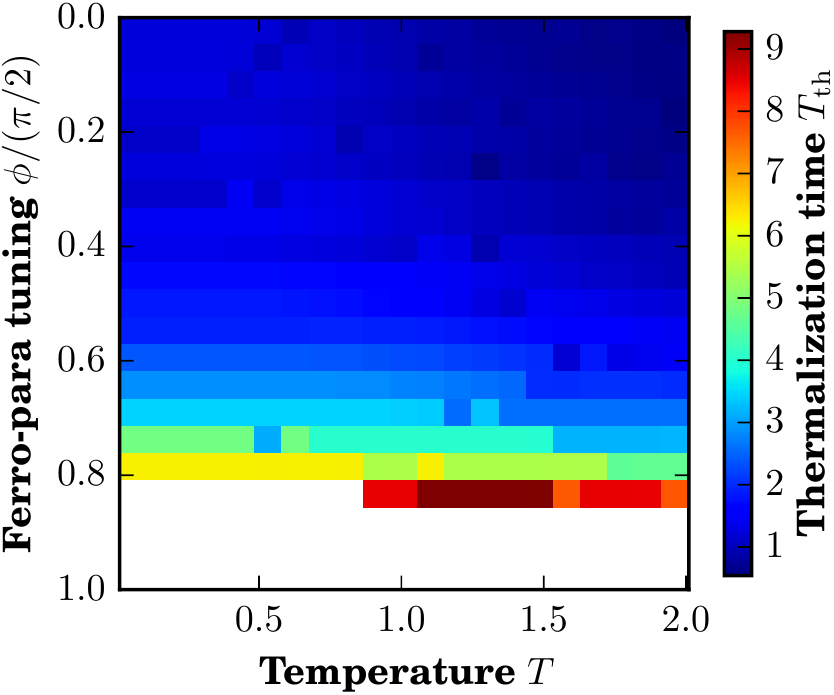}
        \put( 0,85){(a)}
      \end{overpic}
    \end{minipage}\hfill
    \begin{minipage}{0.47\linewidth}
      \begin{overpic}[width=0.9 \columnwidth,unit=1mm]{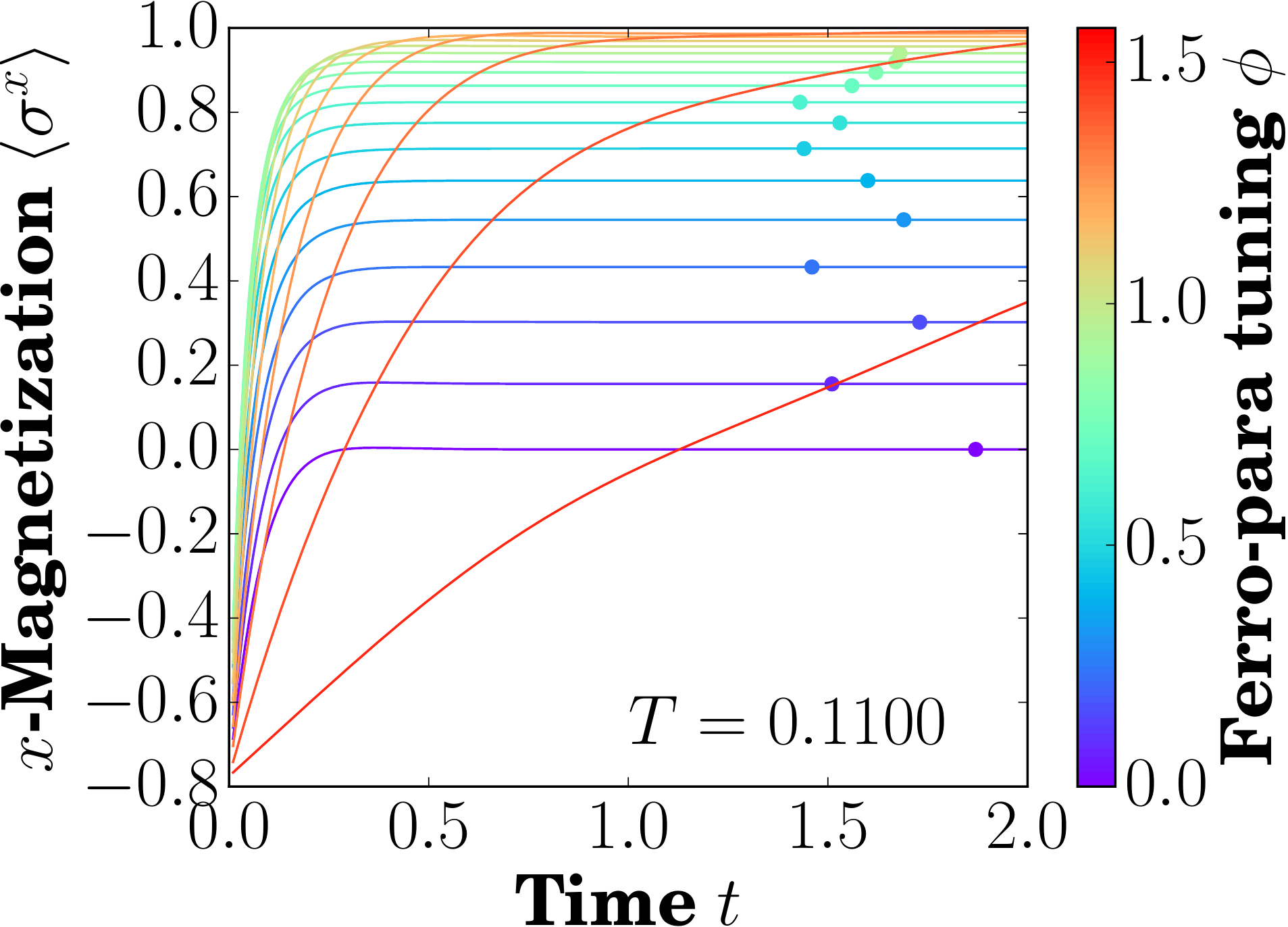}
        \put( 0,80){(b)}
      \end{overpic}
    \end{minipage}
    \caption[Thermalization timescales for the multi-channel Lindblad
      operators with a common bath]
      {\emph{Thermalization timescales for the multi-channel Lindblad
      operators with a common bath.} We consider $N = 2$ qubits and start from an initial random pure state.
      (a)~The thermalization time necessary to approach the steady state value
      of the magnetization according to Eq.~\eqref{eq:tqi:msx}.
      (b)~The local magnetization plotted over time for $T = 0.11$ describes the transient
      behavior to the steady state value. Circles indicate the time
      where the magnetization is first within our tolerance to the steady state
      value.
                                                                                \label{fig:tqi:tscale_L2even}}
  \end{center}
\end{figure}

For a small system of two spins, we analyze the decay by fitting the
average of the magnetization, $\bar{\sigma}^{x}= \frac{1}{N}
\sum_{k=1}^{N} \langle \sigma_{k}^{x} \rangle$ to a double exponential
\begin{eqnarray}                                                                \label{exponentials}
  \bar{\sigma}^{x}= A_1 \exp(\lambda_1 t) + A_2 \exp(\lambda_2 t)
\end{eqnarray}
that captures the main decaying timescales.
For larger systems the number of timescales involved in the evolution
grows exponentially with the number of spins and their contribution
will depend on the initial state, which complicates the design of the
correct fitting function. Therefore, an alternative is to define the
thermalization time as the minimal convergence time to the thermal
value of a reference observable,
\begin{eqnarray}                                                                \label{eq:tqi:msx}
   T_{\mathrm{th}}
  = \min_{t} \left( \left|\bar{\sigma}^{x}(t)
      - \bar{\sigma}_{\mathrm{th}}^{x} \right| < 10^{-10} \right),
\end{eqnarray}
or to the thermal state
\begin{eqnarray}
  T_{\mathrm{th} \mathcal{D}}
  = \min_{t} \left( \mathcal{D}(\rho(t),
        \rho_{\mathrm{th}}^{\mathrm{e}}) < 10^{-10} \right),                 \label{eq:tqi:msxtrd}
\end{eqnarray}
where $\mathcal{D}(\rho, \rho') = \frac{1}{2} \| \rho - \rho' \|$
is the trace distance, $  \rho_{\mathrm{th}}^{\mathrm{e}}$ is defined in
Eq.~\eqref{eq:tqi:thermal_even}, and the corresponding expectation
value of the average magnetization is $\bar{\sigma}_{\mathrm{th}}^{x}$.
We point out that even in the two-qubit case in
Fig.~\ref{fig:tqi:tscale_L2even}, the double-exponential fit cannot resolve
the two timescales over the full range of $\phi$ failing for small $\phi$.
Choosing a relatively tight threshold of $10^{-10}$ ensures capturing the
longest timescales while being definitely above the machine precision of
$10^{-14}$.

The above thermalization timescales can be related to the real part
of the eigenvalues of the Liouville operator. In detail, we can fit
the evolution of the trace distance to an exponential
\begin{eqnarray}                                                                \label{decay}
  \mathcal{D}(\rho(T_{\mathrm{th}\mathcal{D}}), \rho^{\mathrm{e}}_{\mathrm{th}})
  = \mathcal{D}((\rho(0), \rho^{\mathrm{e}}_{\mathrm{th}})
  \exp(-\lambda T_{\mathrm{th}\mathcal{D}}),
\end{eqnarray}
which determines the decaying rate as $\lambda\approx|\Re(\Lambda_j)|$. As
we will show, the value of $j$ of the eigenvalue will depend on the initial
state, which will determine which timescale dominates the decay.
In other words, the decay of different initial states will be governed by
different eigenstates.

% ..............................................................................
\subsection{Two-site systems}
% ..............................................................................

We first consider the case of $N=2$ and examine the multi-channel Lindblad
equation to analyze the common bath \srcasea{} with a coupling
operator $S_{k} = \sigma^{x}$.

Figure~\ref{fig:tqi:tscale_L2even} shows an analysis of the thermalization
time as defined by the convergence of the average mangetization, and
starting from a random pure initial state. Figure~\ref{fig:tqi:tscale_L2even}(a)
shows that the thermalization time in Eq.~\eqref{eq:tqi:msx}
increases when approaching to the paramagnetic phase. Since the average
magnetization is only evolved up to $\tau = 10$, we find a lack of
convergence for $\phi \gtrsim 0.9 \times \pi/2$ as shown by the region
in white color.
The thermalization time $T_{\mathrm{th}\mathcal{D}}$ defined in
Eq.~\eqref{eq:tqi:msxtrd} yields a different result, as it corresponds
to a stricter criterion for thermalization. The maximal difference
between the two definitions for all data points with a thermalization
time smaller than $\tau$ is $3.75$. While the magnetization, e.g.,
$\bar{\sigma}^{x}$ in the $x$-direction from Eq.~\eqref{eq:tqi:msx}, is
more useful with regards to experiments, we will use the trace
distance $\mathcal{D}$ for further calculations.
Figure~\ref{fig:tqi:tscale_L2even}(b) displays the evolution of the
$x$-magnetization at a temperature $T = 0.11$ over time, showing how the
relaxation slows down when approaching the paramagnetic limit.

\begin{figure}[t]
  \begin{center}
    % /data/FinalSpinsInEM3D/02_Thermal_Sizes
    \begin{overpic}[width=0.7 \columnwidth,unit=1mm]{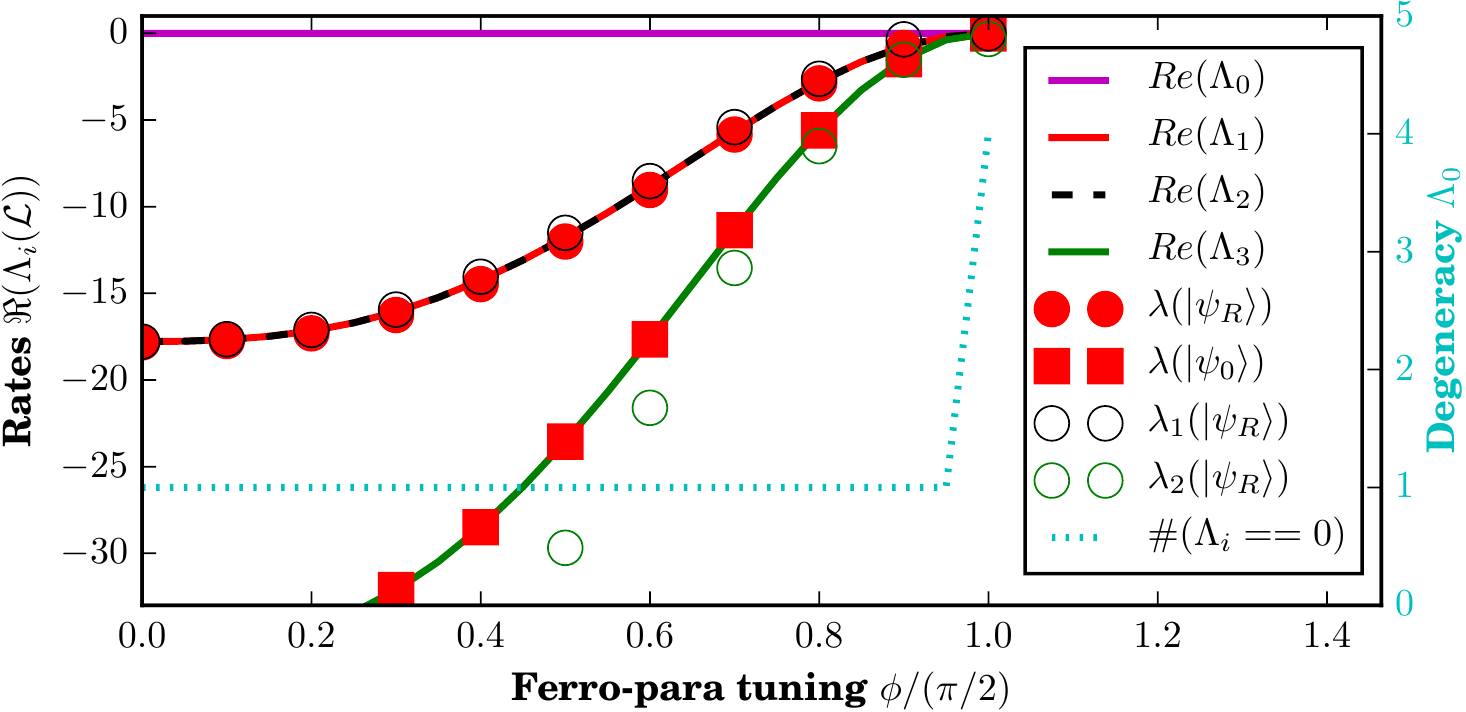}
    %  \put( 0,80){(a)}
    \end{overpic}
    \caption[Spectrum of the Liouville operator for the global
      multi-channel Lindblad operators with a common bath]
      {\emph{Spectrum of the Liouville operator for the global
      multi-channel Lindblad operators with a common bath.} The spectrum
      of the Liouville operator for $N=2$, which defines the
      thermalization timescale.
                                                                                \label{fig:tqi:spectrum_L2even}}
  \end{center}
\end{figure}

Figure~\ref{fig:tqi:spectrum_L2even} displays the analysis of the
eigenstates of the Liouville operator for the same case. Indeed,
all decay rates consistently vanish toward the paramagnetic
limit. For our coupling operator $\sigma^{x}$, this behavior corresponds
to a pure dephasing case where the system does not thermalize. As
a consequence, the gaps between all eigenstates in the
Liouville operator close toward such a limit and the eigenstates
reach full degeneracy.

We now analyze the
Liouville spectrum corresponding to the multi-channel Lindblad equation.
The curves in Fig.~\ref{fig:tqi:spectrum_L2even} show that there is a
single eigenvalue $\Lambda_0$ with a real part that vanishes across the
whole parameter space. This property means that within our symmetry sector
there is a single steady state for the evolution. Furthermore,
the figure displays all other decay rates, which correspond to the
eigenvalues $\Lambda_{j}, j > 0$. We observe that the rates all decay
toward the paramagnetic limit, where a full degeneracy is obtained.
Indeed, this limit corresponds to pure dephasing where the
system does not thermalize.
To analyze the dependency of the decay with the initial states,
we first consider a random initial state $\ket{\psi_{R}}$. The empty
circles show the rates $\lambda_{1,2}$ that result from the fitting
Eq.~\eqref{exponentials} to the evolution of the average
magnetization. The filled circles show the timescale extracted from
the trace distance according to Eq.~\eqref{decay}
and considering the same random initial state $\ket{\psi_{R}}$, while
the filled squares shows the same for an initial  ground state
$\ket{\psi_{0}}$. Notably, the decay of the random initial state
is governed by $\Lambda_1$, while the decay of the ground state does
not depend on the slower timescale and is governed by $\Lambda_2$.

% ..............................................................................
\subsection{Beyond two-site systems}
% ..............................................................................

We now turn to a more complete analysis that goes beyond the two-site
system and consider system sizes up to seven sites in the even symmetry
sector. While still considering a coupling operator $\sigma^{x}$, we
investigate how the system size affects the thermalization timescale.
We investigate how the coupling scheme affects the thermalization. We
question the accuracy of considering neighborhood single-channels
Lindblad operators. In this analysis, we first consider the multi-channel
approach, and then the single-channel approach.

% ..............................................................................
\subsubsection{Thermalization for multi-channel Lindblad operators}
% ..............................................................................

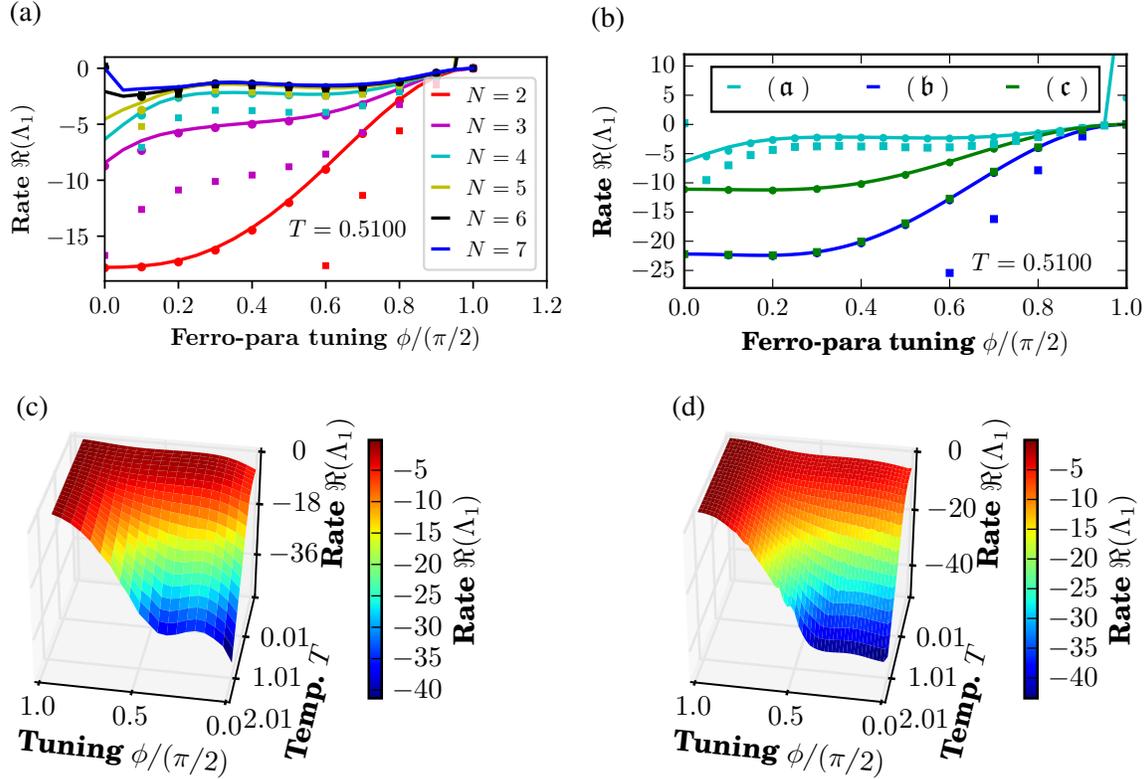
\begin{figure}[t]
  \begin{center}
    \vspace{0.8cm}
    \begin{minipage}{0.47\linewidth}
      % /data/FinalSpinsInEM3D/02_Thermal_Sizes
      \begin{overpic}[width=0.95 \columnwidth,unit=1mm]{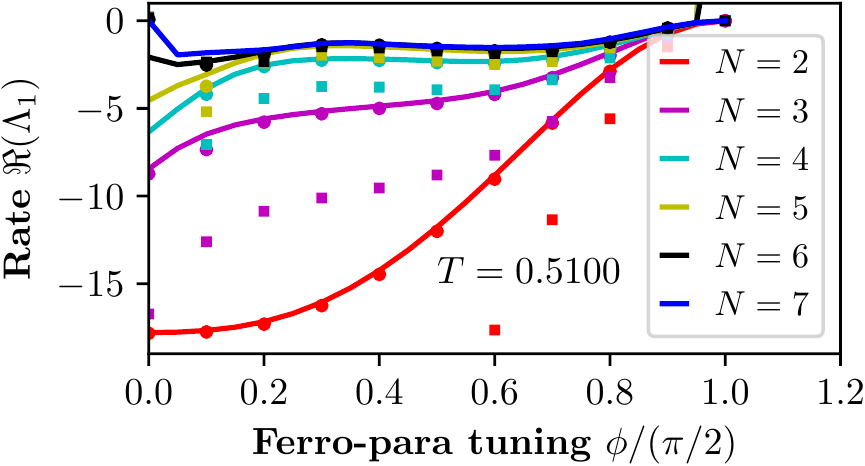}
        \put( 0,60){(a)}
      \end{overpic}
    \end{minipage}%\hfill
    \begin{minipage}{0.47\linewidth}
      % /data/FinalSpinsInEM3D/03_Thermal_Local 03_Thermalization_Local.py
      \begin{overpic}[width=0.95 \columnwidth,unit=1mm]{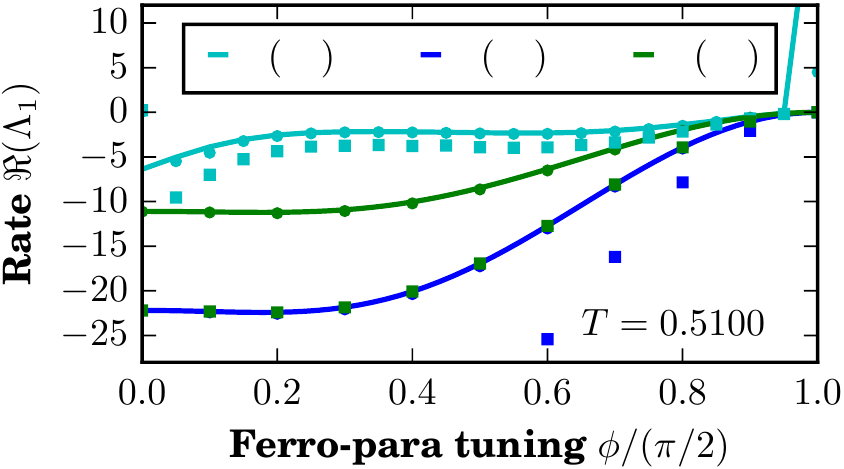}
        \put( 0,60){(b)}
        \put(34,47.5){$\mathfrak{a}$}
        \put(59.5,47.5){$\mathfrak{b}$}
        \put(84.8,47.5){$\mathfrak{c}$}
      \end{overpic}
    \end{minipage}\vspace{0.8cm}
    \begin{minipage}{0.47\linewidth}
      \begin{center}
      % /data/FinalSpinsInEM3D/02_Thermal_Surf
      \begin{overpic}[width=0.8 \columnwidth,unit=1mm]{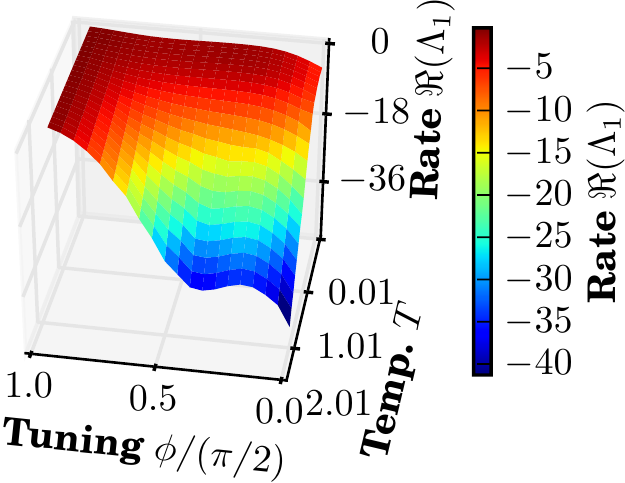}
        \put( 0,78){(c)}
      \end{overpic}\end{center}
    \end{minipage}\hfill
    \begin{minipage}{0.47\linewidth}
      % /data/FinalSpinsInEM3D/03_Thermal_Local_Surf
      \begin{overpic}[width=0.8 \columnwidth,unit=1mm]{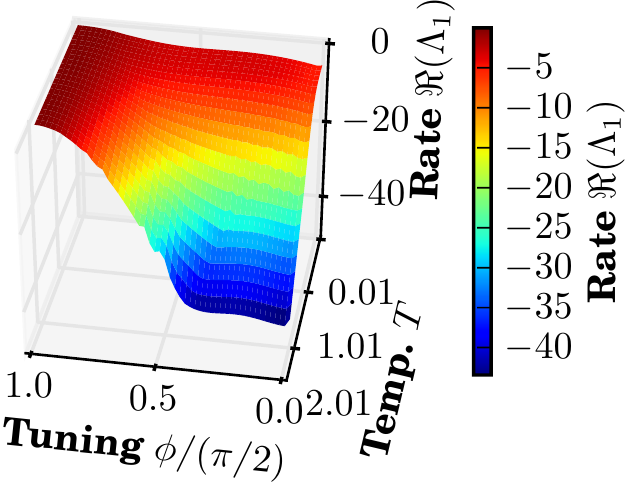}
        \put( 0,78){(d)}
      \end{overpic}
    \end{minipage}
    \caption[Thermalization rates for different systems with multi-channel
      Lindblad operators]
      {\emph{Thermalization rates for different systems with multi-channel
      Lindblad operators.}
      (a)~Decay rates for \srcasea{} for $N = 2$ to $N=7$ spins. Solid curves
      represent the smaller rate $\Re\{\Lambda_1\}$ obtained from the Liouville
      spectrum. Circles (squares) represent the decay rate extracted from the
      trace distance to the thermal state
      starting from a random state (the ground state) in a time evolution. For
      numerical constraints, data for $N = 7$ does not include time evolutions.
      (b)~The rates for different system-bath coupling schemes \srcasea{},
      \srcaseb{}, and \srcasec{} for $N = 4$ spins. Solid curves, circles,
      and squares as in (a).
      (c)~Smaller rate $\Re\{\Delta_1\}$ for $N = 4$ across different
      temperatures $T$ and values for $\phi$ for the common bath \srcasea{}.
      (d)~Smaller rate $\Re\{\Delta_1\}$ for $N = 4$ across different
      temperatures $T$ and values for $\phi$ for independent baths \srcaseb{}.
                                                                                \label{fig:tqi:spectra_even}}
  \end{center}
\end{figure}

Figure~\ref{fig:tqi:spectra_even}(a) shows the smallest values $|\Re(\Lambda_{i})|$,
excluding the steady state,
in the spectrum of the Liouville operator $\mathcal{L}$ of the common
bath for $N \in \{ 2, 3, 4, 5,6,7 \}$. We observe that the gap of the Liouville
operator decreases for increasing system size, leading to slower
thermalization timescales for larger system sizes. Interestingly, we have
observed that this trend also occurs when considering other decay rates
different than the ones given later in the appendix by Eq.~\eqref{eq:tqi:rate_r}, although we do
not present these details here for brevity. As will be further discussed
in Sec.~\ref{sec:fss}, this observation suggests the presence of subradiance in the
emission.

Figure~\ref{fig:tqi:spectra_even}(b) shows the thermalization times for
different system-bath coupling schemes with $N=4$ and $T=0.51$.
The thermalization timescales grow for common, independent, and
end-cap scenarios as one gets closer to the paramagnetic limit,
where the thermalization time approaches infinity. Toward the
ferromagnetic limit the thermalization grows more rapid. Moreover,
the independent and end-cap scenarios \srcaseb{} and
\srcasec{} thermalize faster than with a single common bath
\srcasea{}. Such a common bath introduces spin-spin interactions
mediated by the field which in our case slows down the
thermalization process, and as we will see, even more as the
number of spins in the spin chain grows.
Also, we find that the decay of spins initially in the ground state is
faster than for an initial random state. However, these
differences are more relevant for spins coupled to independent baths
than coupled to a common bath. In addition, when the end-cap configuration
\srcasec{} is initially in the ground state its decay time is exactly the
same as the slowest decay of  \srcaseb{}.

Figures~\ref{fig:tqi:spectra_even}(c) and (d), corresponding to the common
and the independent baths, respectively, show that the qualitative
behavior of the slowest thermalization rate $\Re\{\Lambda_1\}$ with $\phi$ is the
same for all temperatures $T$. In both cases, we observe a faster thermalization
toward the ferromagnetic limit, but the common bath case shows
additionally the presence of a second gap minimum around
$\phi = 0.3 \times \pi / 2$ for high temperatures starting around
$T \ge 0.51$.

The observed features can be experimentally exploited in two possible
directions. On the one hand, a common bath and an initial random
state will slow down dissipation and decoherence processes when
they are not desired, like in certain quantum information schemes.
On the other hand, independent baths and initial ground states can
be used in situations where a thermal state should be reached as fast as
possible. However, we note that the dependence of the thermalization
time on the initial state could be different for low temperatures, since
the ground state is much closer to the thermal state than an excited one.
We dedicate Sec.~\ref{sec:fss} to a finite-size scaling where we estimate
the timescale beyond seven qubits.

% ..............................................................................
\subsubsection{Thermalization for single-channel Lindblad operators             \label{sec:resthermsingleferro}}
% ..............................................................................

To analyze the feasibility of using neighborhood single-channels to simulate
thermalization, we focus on the independent bath configuration
\srcaseb{} and $N=4$. We do not use the $\mathbb{Z}_{2}$ symmetry for
these calculations.

\begin{figure}[t]
  % /data/FinalSpinsInEM3D/04_QuasiLocal_Ferro
  \begin{center}
    \vspace{0.8cm}
    \begin{minipage}{0.47\linewidth}
      \begin{overpic}[width=0.9 \columnwidth,unit=1mm]{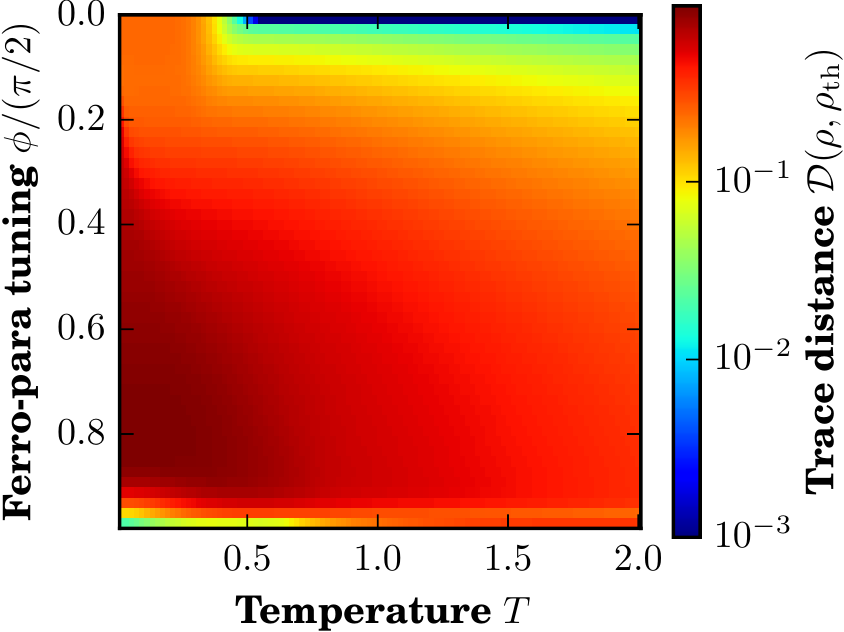}
        \put( 0, 80){(a)}
      \end{overpic}
    \end{minipage}\hfill
    \begin{minipage}{0.47\linewidth}
      \begin{overpic}[width=0.9 \columnwidth,unit=1mm]{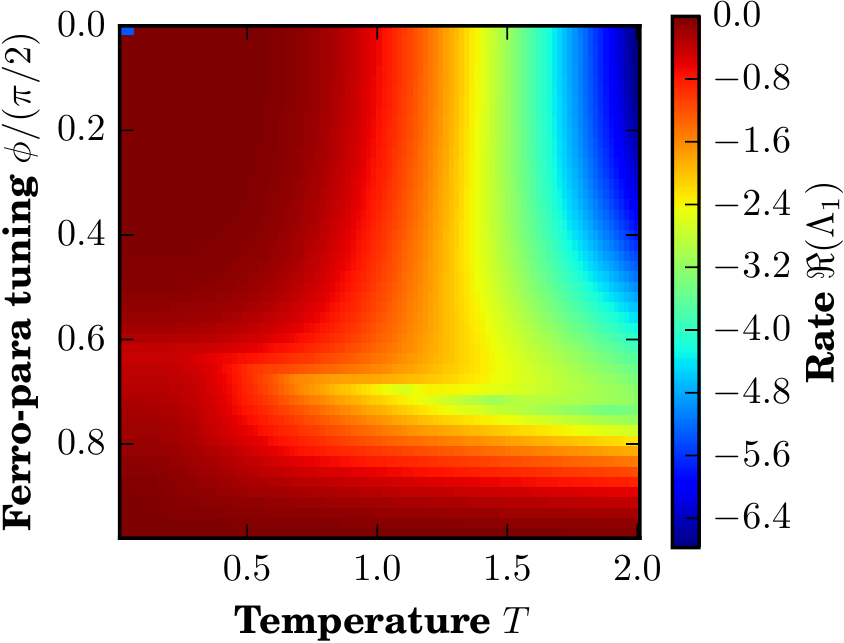}
        \put( 0, 80){(b)}
      \end{overpic}
    \end{minipage}
    \caption[Neighborhood single-channel Lindblad operators]
      {\emph{Neighborhood single-channel Lindblad operators.}
      Three-site Lindblad operators designed for the ferromagnetic limit thermalize
      in scenario \srcaseb{} only in a very small region around the ferromagnetic
      limit, and the multi-channel Lindblad operators are the only way
      to thermalize the system in the complete range of the phase diagram.
      (a)~The trace distance between the time-evolved quantum state and the thermal
      state at the end of the time evolution with $\tau = 10$ shows thermalization
      around the limit $\phi = 0$.
      (b)~The rate given by the eigenvalues of the Liouville operator answers
      the question as to whether the time-evolved state in (a) approaches the thermal
      state at $\tau = 10$ due to long timescales or due to the failure of the
      neighborhood single-channel approach.
                                                                                \label{fig:tqi:fmlimit}}
  \end{center}
\end{figure}

In Fig.~\ref{fig:tqi:fmlimit}(a), we consider the ferromagnetic single-channel
Lindblad operators derived in App.~\ref{neigh_ferro} to evolve a random
initial state until a time $\tau  = 10$ and plot its trace distance
to the thermal state. We set the minimal trace distance to $10^{-3}$
for a meaningful coloring away from the ferromagnetic limit $\phi = 0$. As expected,
the simple single-channel approach makes the system converge to the thermal state in
the ferromagnetic limit, and it does so down to a trace distance of $10^{-14}$.
We observe that we reach relatively small trace distances to the thermal state of
$10^{-2}$ for small perturbations of $\phi \lesssim 0.032 \times\pi/2$
around the ferromagnetic limit. However, the neighborhood single-channel approach
is not valid away from the limiting case. Moreover, we observe that
for temperatures $T \le 0.8$, the system does not thermalize even in the ferromagnetic
limit. The reason for this observation is the trace distance has been calculated at a time
$\tau = 10$ in which the system might not have thermalized yet, predicting
very long thermalization times indeed.

Figure~\ref{fig:tqi:fmlimit}(b) shows that the gap of the single-channel
Liouville operator $\Re(\Lambda_{1})$ decreases for smaller temperatures,
which confirms that a time  $\tau = 10$ is not sufficent to reach the thermal
state in (a) for low temperatures. However, we should stress that the same
argument does imply that, away from the small region around $\phi = 0$,
the system would thermalize when considering longer times. Indeed, the
rate $\Re(\Lambda_{1})$ indicates a fast convergence to a steady state,
but nevertheless this state does not correspond to the thermal one, which
marks the failure of the neighborhood single-channel Lindblad operators beyond
the extreme ferromagnetic limit.

Appendix~\ref{app:tqi:sz} shows numerical results for neighborhood single-channel
Lindblad operators in the paramagnetic limit and we obtain similar results.
In conclusion, the approximation is not very useful for solving
thermalization problems.

% ..............................................................................
\subsection{The thermodyanmic limit via finite-size scaling and subradiance                 \label{sec:fss}}
% ..............................................................................

Finally, we approach larger systems with $N > 7$ via finite size-scaling for
the data points shown in the previous analysis. We observe that the results for
the timescale extracted from the Liouville operator $\mathcal{L}$ converge
for the system sizes of five to seven qubits, i.e., the difference is hardly
visible in the plots. We use a power-law fitting
function of form
\begin{eqnarray}
  \Re(\Lambda_{1,N}) = \Re(\Lambda_{1,N=\infty}) - \frac{c'}{N^{c''}} \, ,
 \label{thermo}
\end{eqnarray}
where $c'$ and $c''$ are positive parameters in the fit that depend on the
parameter $\phi$ within the phase diagram.
This approach is established for the analysis of quantum critical phenomena
in finite-size systems \cite{Cardy2003,SachdevQPT}; the standard
deviation obtained from the fit is a good indicator of whether the
procedure works for the problem given here. The extrapolation for the
thermodynamic limit in Fig.~\ref{fig:tqi:fss_xandz}(a) suggests a non-zero
value for the rate for $N \to\infty$. In addition, it can be seen that
within the standard deviation, the features present in the rate for
seven qubits are also present in our finite-size scaling result. Thus,
our method is promising to analyze the system dynamics for large
open systems beyond the limit of exact diagonalization.

We have previously seen in Fig.~\ref{fig:tqi:spectra_even}(b) that
the decay rate for $N=4$ is smaller for a system coupled to a common
bath than for a system coupled to independent baths. Now, we observe
in Fig.~\ref{fig:tqi:fss_xandz}(a) that for a common bath the decay
rate decreases with N.
Both features suggest the presence of subradiance, a cooperative
phenomenon that leads $N$ spins or atoms to decay much slower than
a single atom \cite{dicke1954}. Classically the subradiance
is described as a result of radiation trapping, i.e., continuous
emission and reabsorption by the atomic sample that leads to a
slowing down of the irreversible energy loss. In the quantum picture,
it is furthermore explained by destructive
interference~\cite{Guerin2016}.

The phenomenon of subradiance was originally described in quantum
optics for independent, i.e., non-correlated atoms; the
decay rate of $N$ atoms coupled to independent
baths is the same one as for a single atom, $\Gamma_N = \Gamma_1$,
while for atoms coupled to a common bath it can either be enhanced
(superradiance) or diminished (subradiance) by a factor dependent on $N$.
In the case of subradiance, the decay time of $N$ atoms $\tau_N$ has
been related to the single atom decay rate $\tau_{at}$ by the
relation $\tau_{N}/\tau_{1}=1+a b_0$, where $a$ is a constant
and $b_0$ is the atomic optical depth \cite{Guerin2016}.

In our case, the presence of correlations in the system presents
two differences with respect to the standard picture.  First,
even for independent baths the decay rate may depend on $N$,
which we do not investigate here. Second, the dependency of the rate on $N$ may be
related to the point of the phase diagram of the open system. Indeed,
we can rewrite  Eq.~\eqref{thermo} as
\begin{eqnarray}
  \frac{\Gamma_N}{\Gamma_1}
  = 1 + \frac{c'}{\Gamma_1} \bigg( \frac{1}{N^{c''}} - 1 \bigg),
\end{eqnarray}
where $\Gamma_1=\Gamma_\infty+c'$. This equation yields a relationship
between decay times, $\tau_{1}/\tau_{N}$ since $\tau_N=1/\Gamma_N$.
Through the dependency with $c'$ and $c''$, we find the
equation that relates the decay time for $N$ spins to that of a
single one depends on the value of $\phi$, and therefore the amount
of correlations within the spins.
While few works describe collective effects for atoms that are coupled
through a hopping term \cite{Asenjo2017, Plankensteiner2018,Facchinetti2016},
even fewer studies describe the effect of atom correlations, like for
instance in the context of one-dimensional waveguides \cite{Ruostekoski2017}
or considering just two atoms \cite{Fiscelli2018}. Therefore, the interplay
between strong correlations and collective effects in the dynamics of many
body systems is still a largely unexplored domain.

% ..............................................................................
\subsection{Two coupling operators in the interaction Hamiltonian}              \label{beyond}
% ..............................................................................

\begin{figure}[t]
  % /data/FinalSpinsInEM3D/01_Thermal_Scales
  \begin{center}
    \vspace{0.5cm}
    \begin{minipage}{0.47\linewidth}
      \begin{overpic}[width=0.9 \columnwidth,unit=1mm]{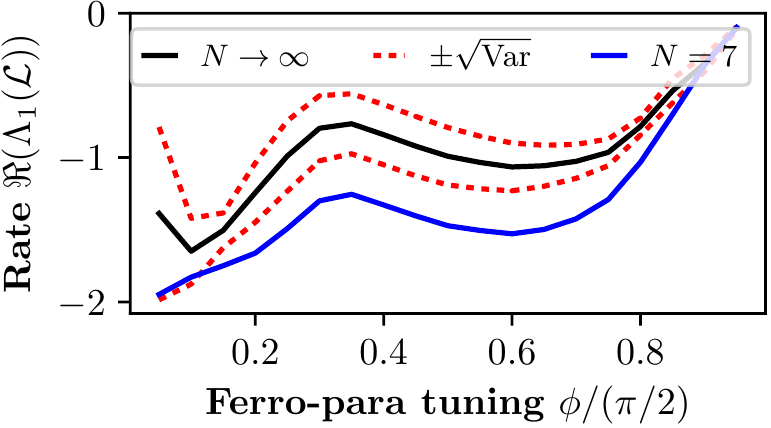}
        \put( 0,60){(a)}
      \end{overpic}
    \end{minipage}\hfill
    \begin{minipage}{0.47\linewidth}
      \begin{overpic}[width=0.9 \columnwidth,unit=1mm]{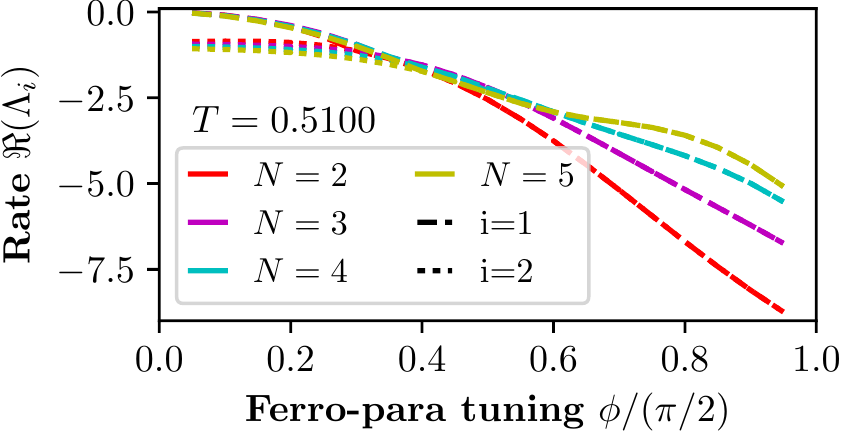}
        \put( 0,60){(b)}
      \end{overpic}
    \end{minipage}
    \caption[Additional aspects of the open system quantum Ising model.]
      {\emph{Subradiance and generalized coupling.}
      (a)~Finite-size scaling extrapolates the gap of the Liouville operator
      for system sizes $N \to \infty$. This approach predicts an open gap
      of the Liouville operator $\mathcal{L}$ toward the
      thermodynamic limit (including taking the variance
      of the fit into consideration). For comparison, we show the
      result for $N = 7$.  We hypothesize a subradiance effect.
      (b)~Treating a final bath case $\msrcased$, we find for multiple
      system-bath coupling operators $\sigma^{x}+\sigma^{z}$ in the
      interaction Hamiltonian the system thermalizes across the complete
      phase diagram except in the limit $\phi =0$.  Thermalization
      times are deduced from the eigenvalues $\Re(\Lambda_{1})$ and
      $\Re(\Lambda_{2})$ of $\mathcal{L}$. System sizes are color coded.
      First ($i=1$, dot-dashed) and second ($i=2$, dotted) non-zero
      eigenvalues are indicated by line style. Curves may overlap.
                                                                                \label{fig:tqi:fss_xandz}}
  \end{center}
\end{figure}

We have seen in the previous sections that the relaxation dynamics is
highly sensitive to the parameters and structure of both system and
interaction Hamiltonians. To illustrate this observation further, we end our analysis
by extending beyond our previous common bath
scenario. We take the interaction Hamiltonian to include not just one but
two coupling operators, namely $\sigma_k^{x}$ and $\sigma_k^{z}$, such that we
have a final scenario referred as $\msrcased$,
\begin{eqnarray}
  H^{\msrcased}_{\mathrm{S+R}}&=& H_{\mathrm{z,S+R}}+H_{\mathrm{x,S+R}}=%\cr
  \sum_{k=1}^{N} \sum_{\bf q} \sum_{\alpha \in \{x, z \}}
    \sigma_{k}^{\alpha} (g_{k{\bf q}} b_{{\bf q}} + g_{k{\bf q}}^{\ast} b_{{\bf q}}^{\dagger}) \, .
\end{eqnarray}
Figure~\ref{fig:tqi:fss_xandz}(b) contains the first two eigenvalues of
the Liouvillian for two to five qubits. Since the $\mathbb{Z}_{2}$ symmetry
is broken by the operator $\sigma_k^{z}$ in the interaction Hamiltonian,
thermalization occurs in the whole Hilbert space for every parameter
regime. An exception is the ferromagnetic limit ($\phi=0$), where the gap
for the first eigenvalue closes.
In the ferromagnetic limit, the $\sigma_k^z$ part in the interaction Hamiltonian is responsible for breaking
the symmetry; this part commutes with the system Hamiltonian. It follows that the terms does
not produce any transition between eigenstates with different energies,
using moreover the properties of the Pauli matrices.
It contributes in the Lindblad equation with a term
proportional to $\gamma(E_{ba}=0, {\bf r}_{jk})$. However, as discussed
in App.~\ref{sec:tqi:full}, $\gamma(0, {\bf r}_{jk})=0$ in our case.
Thus in this limit, the term $H_{\mathrm{z,S+R}}$ does not contribute
at all to the dynamics. In other words, the system evolution is determined
only by $H_{\mathrm{x,S+R}}$, which preserves the symmetry. This reasoning explains the
double degeneracy of the ground states of the Liouville operator (with
two eigenvalues $\Re\{\Lambda_0\}=\Re\{\Lambda_1\}=0$), while the finite
non-zero value of $\Re\{\Lambda_2\}$ leads to the conclusion that each
symmetry sector thermalizes.

% ..............................................................................
\subsection{Experimental realizations}                                          \label{experiment}
% ..............................................................................

%
%
Quantum simulator platforms realizing the quantum Ising model
include ultracold atoms and molecules \cite{Carr2011,carr2013j}, trapped ions
\cite{Bohnet2016}, and Rydberg systems \cite{Labuhn2016}. Modelling the
bath coupled to these architectures or even understanding the accuracy
of the Lindblad description are difficult problems on their own. Here
we have considered for simplicity a bath consisting of a
three-dimensional electromagnetic field present in AMO-based (atomic, molecular,
and optical) quantum simulator platforms, which couples to the spins of our
system via dipole interactions. The black-body radiation of the
surrounding experiment can serve as motivation to choose such a field.
In addition, trapped ions \cite{Schmied2008}, Rydberg atoms
\cite{Everitt2009} and ultracold atoms \cite{Zoubi2013} can be confined in an
optical cavity considered in the bad cavity limit, where the electromagnetic
field has a broad-band density of states and therefore produces the
desired Markovian interaction \cite{Mateus2004}.

Although we use the
electromagnetic field as a basis throughout the paper, a set of
interacting spins coupled to a continuous bosonic bath can be
implemented with other models. One possibility is to consider ultracold
atoms coupled to the bosonic field formed by the excitations of a
Bose-Einstein condensate (BEC) as proposed in \cite{Recati2005}
and extended in \cite{Orth2008} for the Ising model. A close
alternative is to consider atoms with two internal levels. The first
level is trapped by an optical lattice and implements the
open system dynamics.  The second level is untrapped and plays
the role of a bath \cite{deVega2008,Stewart2017}. Similarly,
recent progress has enabled the coupling of qubits to acoustic phonons
in superconducting qubits \cite{Chu2017}, although the long coherence
times between qubit and phonon modes cannot be accurately described
with our approach based on the Born-Markov approximations. However,
both BEC excitations at low momenta \cite{Marino2017} and acoustic
phonons have a linear dispersion, analogous to photons in the
electromagnetic field. This analogy draws the similarity to the model here
studied and further emphasizes the experimental applicability of our work.

% =============================================================================
\section{Conclusions                                                           \label{sec:tqi:concl}}
% =============================================================================

% Approach
% --------

We have analyzed the dynamics of a quantum Ising model coupled to
a thermal field in the Markov limit by considering a multi-channel
Lindblad equation, in particular thermalization and decoherence times.
Correct calculation of decoherence times is vital to accuracy and
verifiability of any quantum simulator, including the many experimental
examples treating the quantum Ising model. Such a multi-channel Lindblad
equation satisfies the Born-Markov and secular approximations well known
to Lindblad approaches and makes no further approximation.  In contrast,
we have found that the commonly used single-channel or local
approximation produces no thermalization in the simulations for
the quantum Ising model; thermalization timescales cannot be predicted
with this approach.

% Results
% -------

We explored coupling with the bath in three different experimental
scenarios, i.e., common, independent, and end-cap baths. We also explored
how thermalization times depend on both the phase diagram and the number
of spins in our system. We showed how the bath, system size, and region
in the phase diagram can work together to speed up or slow down
thermalization, which can be used to protect many-body states from
decoherence. For instance, one could quench the system into a
system-bath-phase region protected from decoherence and reverse the
quench to regain the state when needed. The errors introduced due to
non-adiabatic quenches might be less critical than the errors
from decoherence while storing the state for a long time. We emphasize
that very little is known about control of many-body decoherence, with
most decoherence times given in terms of one or a few qubits subject
to gate operations in a ``digital'' quantum computer, and scaling to
larger systems assumed, a well-known problem in trapped ions, for
example. In contrast, our study applies directly to calculations on
``analog'' quantum computers, i.e., quantum simulators, as well as to
adiabatic quantum machines like D-Wave.  Our study can also shed
light on the coherence times of large many-body states in digital
quantum computers between gate operations.
\vspace*{3mm}

In detail, we found the following.
\begin{tasks}
\task The thermalization depends on the phase of the Ising model and
  on whether the $\mathbb{Z}_{2}$ symmetry is preserved. For a system
  coupling operator $\sigma^{x}$, the symmetry is preserved and
  thermalization is produced in each  symmetry sector, except
  in the paramagnetic limit where the interaction produces pure
  dephasing. Moreover, thermalization times are found to increase
  when approaching this limit for any configuration and spin
  number. For a system coupling operator $\sigma^{x}+\sigma^{z}$
  (or simply $\sigma^{z}$) the symmetry is broken and the system
  thermalizes in the full Hilbert space. In the ferromagnetic limit,
  the symmetry-breaking component of the coupling, $\sigma^{z}$, has
  virtually no effect and the system thermalizes in each sector due
  to the action of $\sigma^{x}$.
\task Spins coupled to a common bath thermalize slower than spins
  coupled to independent baths, and the thermalization time increases
  with the number of spins and also depends on the system phase. This result
  suggest an interplay between collective effects leading in this
  case to subradiance and system correlations. Independent baths
  thermalize faster if all sites are coupled to a bath instead of
  only one site, but in both cases the thermalization time is highly
  dependent on the initial state. These results are supported by data
  from the spectrum of the Liouville operator and the time evolution
  of the Lindblad equation. Thus, we suggest the independent
  bath scenario to enhance thermalization or a common bath
  to suppress it and extend coherence times.
\task The thermalization time of atoms is structured in $\phi$. For
  instance, for atoms coupled to a common reservoir the timescale
  shows a relative maximum around $0.3 \times \pi / 2$ at $T=0.51$.
  The presence of such maxima can be used as \textit{sweet spots}
  to preserve coherence in the system away from the values of
  $\phi$ close to the ferromagnetic case.
\task The proposed neighborhood single-channel Lindblad operators
  designed for the ferromagnetic and paramagnetic limits gives rise
  to a thermal steady state only in a very narrow region around
  the extreme limits. However, even in these cases, it does not
  reproduce the correct decay rate given by the multi-channel
  Lindblad equation.  Thus, the computationally intensive
  multi-channel Lindblad approach appears to be necessary to
  estimate many-body thermalization and decoherence times.
\end{tasks}

% Outlook
% -------

% Options within our model / modifying the open system
Our approach provides a clear path forward toward addressing
pressing open problems suitable for near-term research projects,
both within the present classical computational limits of 6-7 qubits
imposed by the necessary multi-channel Lindblad approach, as well
as the much larger system size that can be treated in experimental
quantum simulator architectures.
For instance, one may explore thermalization and decoherence control
by varying the angle between the dipoles in the bath and the spins
in the system in a common vs. independent reservoir scenario. In the
latter, one could treat varying the dipole direction across the
independent baths.  The distance between the spins could
also be changed to a non-equidistant spacing to
reproduces what happens, for example, in one-dimensional
ion traps. There are other questions which can be addressed within this
framework, such as coupling to baths at different temperatures and
the resulting equilibrium temperature and possible temperature gradients.
Another extension to this research is to include long-range interactions
in the system Hamiltonian. These long-range interactions present a
correction to local tight-binding models such as the transverse
quantum Ising model, occur in a number of experimental platforms, and
can have significant effects on the statics and dynamics of a closed system
\cite{JaschkeLRQIC}.  Could long-range effects in an open quantum system
be used to control thermalization and decoherence times, and how would
such control interact with different bath scenarios as in this study?
%

% Relaxing approximations
The validity of the secular approximation in many-body systems should
also be revisited; the large number of possible transitions will increase the
possibility of smaller differences between two transition frequencies for
an increasing number of spins. Moreover, it remains an open question if
the conclusions are altered when going beyond the Markov and weak coupling
approximations. This question relates to whether thermalization is present in
non-Markovian open quantum systems \cite{Rivas2014,Breuer2016,deVega2017}.
%

% Other types of baths
Relaxing these approximations can open the way to consider further
experimental setups which are not covered by the approximations of
the Lindblad equation. An example is a sympathetic
cooling in ultracold atomic and molecular experiments~\cite{carr2004a,Lim2015}.
Ultracold molecules have been considered in the
context of collisions leading to decoherence of the
system~\cite{Hemming2010,Braaten2017}, which can be described with the Lindblad
master equation. However, considering the bath as a quantum gas
may allow one to explore Fermi vs. Bose statistics, by selecting the
corresponding isotopes, or choosing a coupling which ranges from weakly
to strongly interacting. Furthermore, some open systems cannot be
described with a Markov approximation such as ladder-type structures
where the system is of the same size as the bath \cite{Kaufman2016},
i.e., one rail representing the system, and the other
the bath. Another example is that of a few nuclear spins of $^{13}$C
atoms surrounding a color center \cite{Smeltzer2011}.
% Other Markovian baths?
% Anharmonic baths.
%
% Extending results toward the thermodynamic limit
Future research may focus on local or neighborhood single-channel
operators leading to the thermalization of the system, for and beyond the
quantum Ising model. Such a step would help to move beyond seven spins
and understand further the thermodynamic limit.

\emph{Acknowledgments $-$}
We gratefully appreciate discussions with K.~Maeda, S.~Montangero,
G.~Shchedrin, and N.~C.~Smith. This work was performed in part at the Aspen
Center for Physics, which is supported by National Science Foundation
grant PHY-1607611. This work was performed with partial support of the NSF
under grants PHY-1806372, OAC-1740130, CCF-1839232, and the AFOSR under
grant FA9550-14-1-0287. We acknowledge support of the U.K. Engineering and
Physical Sciences Research Council (EPSRC) through the ``Quantum Science
with Ultracold Molecules'' Programme (Grant No. EP/P01058X/1). The
calculations were carried out using the high performance computing
resources provided by the Golden Energy Computing Organization at the
Colorado School of Mines. I.D.V. was financially supported by the
Nanosystems Initiative Munich (NIM) under project No. 862050-2 and
the DFG-grant GZ: VE 993/1-1.

% =============================================================================
% Bibliography
% =============================================================================

%\bibliography{refs}
%\bibliographystyle{unsrt}

%merlin.mbs apsrev4-1.bst 2010-07-25 4.21a (PWD, AO, DPC) hacked
%Control: key (0)
%Control: author (0) dotless jnrlst
%Control: editor formatted (1) identically to author
%Control: production of article title (0) allowed
%Control: page (1) range
%Control: year (0) verbatim
%Control: production of eprint (0) enabled
%

\appendix

% =============================================================================
\section{Thermalization for odd symmetry sector and $S_{k} = \sigma_{k}^{z}$    \label{app:tqi:odd}}
% =============================================================================

\begin{figure}[htbp]
  \begin{center}
    \vspace{0.8cm}
    \begin{minipage}{0.4\linewidth}
      % /data/FinalSpinsInEM3D/11_Thermal_Scales
      \begin{overpic}[width=0.95 \columnwidth,unit=1mm]{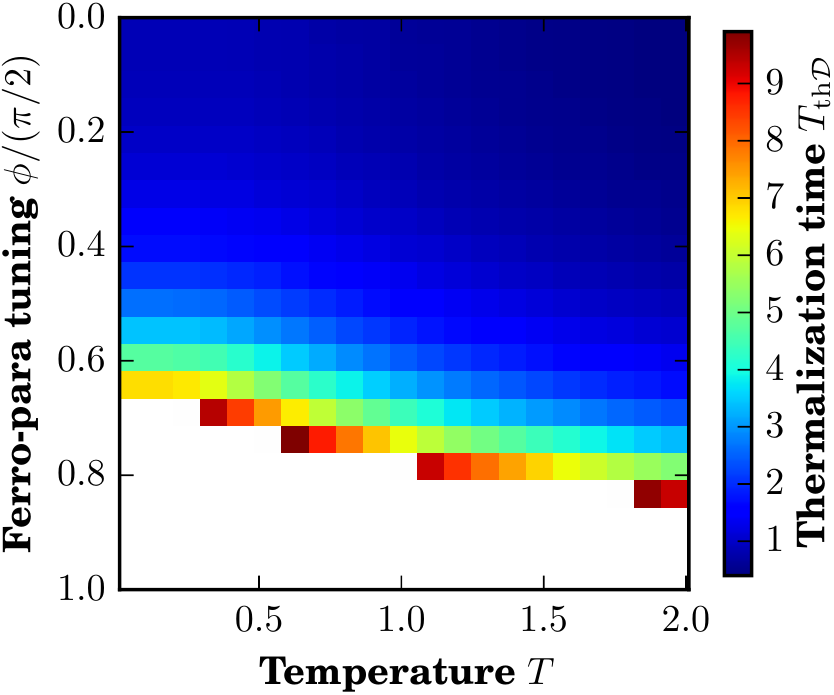}
        \put( 0,87){(a)}
      \end{overpic}
    \end{minipage}\hfill
    \begin{minipage}{0.58\linewidth}
      \begin{center}
      % /data/FinalSpinsInEM3D/12_Thermal_Sizes
      \begin{overpic}[width=0.95 \columnwidth,unit=1mm]{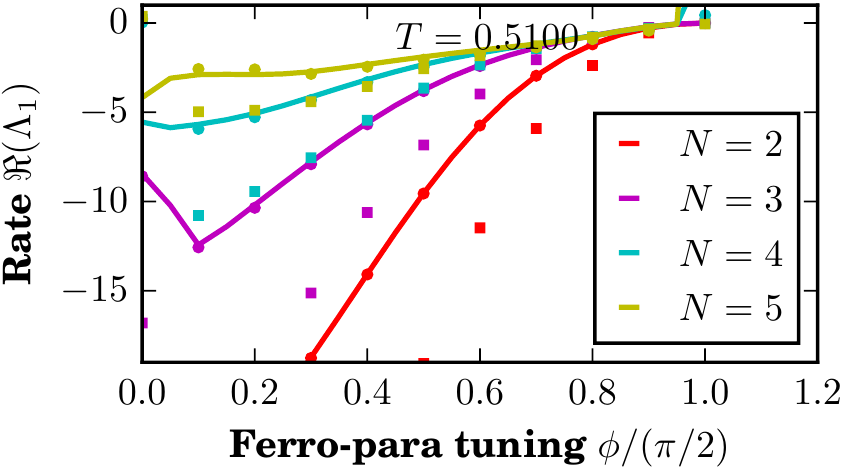}
        \put( 0,59){(b)}
      \end{overpic}\end{center}
    \end{minipage}\hfill
    \caption[Summary for the odd symmetry sector]
      {\emph{Summary for the odd symmetry sector.}
    (a)~Thermalization timescale for the odd symmetry sector via the
    trace distance. We observe the same trend as in the even sector, i.e.,
    the timescale becomes longer toward the paramagnetic limit. Time
    evolution for $\tau = 10$; white regions do not meet criterion within
    this time.
    (b)~Rate and timescales for different system sizes for the global
    multi-channel approach and initial states. The curves represent the
    closing rate toward the paramagnetic limit and for growing system
    size. A random initial state (circles) matches this rate of the
    Liouville operator $\mathcal{L}$; the ground state (squares) as the
    initial state thermalizes faster.
                                                                                \label{fig:tqi:fmodd}}
  \end{center}
\end{figure}

So far, we have considered the analysis of the system in the even symmetry sector of the
$\mathbb{Z}_{2}$ symmetry. We summarize the odd sector in this brief
appendix. We expect similar behavior. The thermalization time
$T_{\mathrm{th}\mathcal{D}}$ slows down toward the paramagnetic limit,
see Fig.~\ref{fig:tqi:fmodd}(a) for multi-channel Lindblad operators in
scenario \srcasea{}. If we consider increasing system size $N$ within the
same setup, we note that the gap is closing for an increasing number of
spins. The initial state has an equal effect on the timescale as in the
even sector: the ground state (squares) has a shorter thermalization time
than the random initial state (circles); the latter match the gap
$\Re(\Lambda_1)$ in the spectrum of the Liouville operator $\mathcal{L}$.
We conclude from these two examples that the features presented in our
work are not unique to the choice of the symmetry sector.

% =============================================================================
\section{Multi-channel Lindblad equation\label{sec:tqi:full}}
% =============================================================================

In this appendix, we present a more detailed description of the multi-channel
approach. We will specify the component $\gamma(E_{ba}, {\bf r}_{jk})$
present in the decay rates  of the Lindblad equation, see
Eq.~\eqref{eq:tqi:ratederiv2}. Using standard notation, we
denote the transition frequency
$\omega =E_{ba}$. As discussed in \cite{Lehmberg1970}, for the light-matter
interaction the function $\gamma(E_{ba}, {\bf r}_{jk})$ takes the form
\begin{eqnarray}                                                               \label{eq:tqi:rate_r0}
  \gamma(\omega, 0)
  = \tilde{n}(\omega) \underbrace{\frac{4 \omega^3 d_{\mathrm{dip}}^2}{3 c^3}}_{= \Gamma_0}
  = \tilde{n}(\omega) \Gamma_0 \,,
\end{eqnarray}
for the case $j = k$, where $c$ is the speed of light. For non-zero
spatial distance, it is found that
\begin{eqnarray}                                                               \label{eq:tqi:rate_r}
  \gamma(\omega, r_{jk})
  &=& \tilde{n}(\omega) \Gamma_0 {\Re}\big(\chi(\omega,{\bf r}_{jk})\big),
\end{eqnarray}
where
\begin{eqnarray}                                                                \label{chi}
\chi(\omega,{\bf r}_{jk})=\Bigg((1 - \xi^2)
  \frac{\mathrm{-i} \mathrm{e}^{\mathrm{i} \omega r_{jk}}}{\omega r_{jk}}
  + (1 - 3 \xi^2) \left(
  \frac{\mathrm{e}^{\mathrm{i} \omega r_{jk}}}{(\omega r_{jk})^2}
  + \frac{\mathrm{i} \mathrm{e}^{\mathrm{i} \omega r_{jk}}}{(\omega r_{jk})^3}
  \right) \Bigg) \, .
\end{eqnarray}
Here we assume for simplicity that the spin dipoles in the bath are
aligned orthogonally to the Ising chain, which leads to
$\xi \equiv (\vec{d}_{\mathrm{dip}} \cdot \vec{r}_{jk}) / (d_{\mathrm{dip}} r_{jk})
= 0$, with $r_{ij}=|{\bf r}_{ij}|$.
We note that different choices of dipole orientation lead to different
spatial decays for spin-spin correlations within the dissipation
process. For instance, if the dipoles and spin locations are parallel,
we find $\xi = 1$, and therefore the first term in
Eq.~\eqref{eq:tqi:rate_r} cancels while the second and third order term
in $1 / r_{jk}$ survive. This would lead to spin-spin correlations
that decay faster with their distance than in cases where the first
slower decaying term is present. In contrast, for $\xi = 1 / \sqrt{3}$,
the second and third order terms in $1 / r_{jk}$ is set to zero, while
the slower decaying term remains with a pre-factor of $2 / 3$. This case
is apart from the pre-factors such as $\Gamma_{0}$, as in the case
of spins coupled to a BEC, and shows the versatility with
which this system can be tuned. We ran simulations
similar to Fig.~\ref{fig:tqi:spectra_even}(a) for selected values of $\xi$
and observed the same behavior, i.e., the thermalization timescale
slows down for more spins. Furthermore, we have defined
$\tilde{n}(\omega)$ as
\begin{eqnarray}                                                               \label{eq:tqi:bestat}
  \tilde{n}(\omega) &=& \begin{cases}
  n(\omega) \, , \quad \forall \omega > 0 \, , \\
  1 + n(-\omega) \, , \quad \forall \omega < 0 \, .
  \end{cases}
\end{eqnarray}
where $n(\omega)=1/(e^{\beta\omega-1}-1)$, and $\beta=1/k_\mathrm{B}T$,
with $k_\mathrm{B}$ the Boltzmann constant. Finally, we neglect for
simplicity the Lamb shift-like terms that come from the imaginary part
of $\chi(\omega,{\bf r}_{jk})$, since these terms will not contribute to the
dissipative timescales and the thermalization process we aim to
analyze here.

% =============================================================================
\section{Deriving the neighborhood single-channel Lindblad operators           \label{sec:tqi:limits}}
% =============================================================================

We introduce the neighborhood single-channel Lindblad
operators in this section.

% -----------------------------------------------------------------------------
\subsection{Neighborhood single-channel operators in the paramagnetic limit}
% -----------------------------------------------------------------------------

To derive the simpler Lindblad operators that in the paramagnetic limit give
rise to thermalization, we consider solely for this subsection
$S_{k} = \sigma_{k}^{z}$. For the paramagnetic limit the eigenstates
are product states of local eigenstates of the Pauli matrices, i.e.,
product states of spins aligned and anti-aligned with the external field.
The Schmidt decomposition of a product state has exactly one non-zero
singular value, e.g., $\lambda_{1ka} = 1$, and therefore
Eq.~\eqref{eq:tqi:schmidtfull} reduces drastically such that
\begin{eqnarray}                                                                \label{eq:tqi:condIandII}
  ^{ij'}\mathcal{F}_{ab}^{[k]}
  &=& \underbrace{\bra{a_{k}^{i'=1}} S_{k} \ket{b_{k}^{j=1}}}_{(I)} \; %\cdot
  \underbrace{\braket{a_{\bar{k}}^{i'=1}}{b_{\bar{k}}^{j=1}}}_{(II)} \, , \lambda_1 =
  \lambda_{1}' = 1 \, ,
\end{eqnarray}
with all other Schmidt values being zero. In order to contribute,
both terms $(I)$ and $(II)$ should be non-zero. Moreover, since
we use an orthogonal set of basis vectors for $k$ and $\bar{k}$
each of the terms is either $0$ or $1$.
The term $(I)$ marked in Eq.~\eqref{eq:tqi:condIandII} is
equal to one if $\bra{a_{k}^{i'=1}}$ and $\ket{b_{k}^{j=1}}$ are
anti-aligned as $S_{k}$ flips the spin. The term $(II)$ from
Eq.~\eqref{eq:tqi:condIandII} is equal to one if the sites
$k' \neq k$ do not change their state. Thus, we obtain that
the Lindblad operator for each site, producing cooling of the
system is
\begin{eqnarray}                                                                \label{eq:tqi:paracool}
  L^{[k]} &=& \ket{\uparrow_{x}}_k \bra{\downarrow_{x}}_{k} \, ,
\end{eqnarray}
while $(L^{[k]})^\dagger$ produces heating, respectively.
The state on site $k$ aligned with the external field in the
$x$-direction is $\ket{\uparrow_{x}}_k$, while the state
anti-aligned is $\ket{\downarrow_{x}}_k$. Therefore, the
Lindblad operator in Eq.~\eqref{eq:tqi:paracool} aligns
the spin $k$ with the external field and brings this spin
to the ground state of the paramagnetic limit.
A numerical implementation of this case is discussed in App.~\ref{app:tqi:sz}.

% -----------------------------------------------------------------------------
\subsection{Neighborhood single-channel operators in the ferromagnetic limit}  \label{neigh_ferro}
% -----------------------------------------------------------------------------

To develop the single-channel operators for the ferromagnetic limit, we
return to the coupling $S_{k} = \sigma_{k}^{x}$.  On the one hand, we know
that the eigenstates are not entangled, i.e., $\lambda_j=0$ for $j\neq 1$
but degenerate. For instance, the ground state subspace contains two
degenerate, symmetry-broken eigenstates, $|\uparrow\cdots\uparrow\rangle$ and
$|\downarrow\cdots\downarrow\rangle$. Thus, we do not use the
$\mathbb{Z}_{2}$ symmetry for this setup. We neglect for simplicity
the edges of the chain. The first excited subspace corresponds to
a set of $2 (N-2)$ degenerate states with an energy $4$ with respect
to the ground state. These are the excited states which can be reached after an action of a single
Lindblad operator from one of the ground states. The states contain two domain walls which separate
a phase of spins $\uparrow$ (or $\downarrow$) from a background phase of
spins oriented in the opposite direction. In a similar way, the second
excited subspace will be composed of states with four domain walls, and
an energy $8$ with respect to the ground state. In general, the different
energy subspaces will correspond to eigenstates having an increasing
number of magnetic domains.

On the other hand, from the spectral decomposition of the coupling
operator in Eq.~\eqref{eq:tqi:schmidtfull} and the multi-channel
Lindblad equation, we know that (i) $\sigma_{k}^{x}$ only connects
eigenstates separated by a single spin flip, and (ii) in the Lindblad
equation only transitions with $\gamma(\omega\neq 0,{\bf r}_{jk})$ will
contribute, since $\gamma(0,{\bf r}_{jk})=0$. In other words, Lindblad
operators that do not change the energy will not be taken into account.
Hence, both conditions imply  that Lindblad operators should connect
eigenstates that differ in a single spin flip and contain a different
number of magnetic domains, i.e., correspond to a different energy.

With this prescription, the simplest choice are Lindblad operators
defined on a neighborhood of up to three sites.
Depending on whether they increase or decrease the number of
magnetic domains, they will be associated with heating or cooling.
In detail, we define the states
$\ket{m}_{k} = \statem{k}$,
%$\ket{n}_{k} = \staten{k}$,
$\ket{o}_{k} = \stateo{k}$,
%$\ket{p}_{k} = \statep{k}$,
%$\ket{q}_{k} = \stateq{k}$,
$\ket{r}_{k} = \stater{k}$, and
%$\ket{s}_{k} = \states{k}$, and
$\ket{t}_{k} = \statet{k}$ on the three
neighboring sites $k - 1, k, k + 1$. We can now define a generic Lindblad
operator $L_{ab}^{[k]} = \ket{a}_{k} \bra{b}_{k}$ acting on the bulk of the
system on a three-site neighborhood. In detail, the heating Lindblad
operators have the form
\begin{eqnarray}                                                                \label{eq:tqi:Lferroheat}
  L_{rt}^{[k]} &=& \ket{r}_{k} \bra{t}_{k} \, , \,\,
  L_{om}^{[k]} = \ket{o}_{k} \bra{m}_{k} \, ,
\end{eqnarray}
for $k = 2, \ldots, N - 1$, and give rise to an increase in energy.
Each of these Lindblad operators can be interpreted as a
spin flip with one or two control-spins. Furthermore, for
the boundary sites, we define two-site states $\ket{u}_{k} = \stateu{k}$,
$\ket{v}_{k} = \statev{k}$, $\ket{w}_{k} = \statew{k}$, and
$\ket{x} = \statex{k}$ for the sites $k$ and $k + 1$. The
corresponding Lindblad operators are $L_{wx}^{[1]}$,
$L_{vu}^{[1]}$, $L_{vx}^{[N-1]}$, and $L_{wu}^{[N-1]}$ for the boundary
sites.
The cooling operators then correspond to their Hermitian
conjugates, i.e.,
\begin{eqnarray}                                                                \label{eq:tqi:Lferrocool}
  L_{tr}^{[k]} &=& \ket{t}_{k} \bra{r}_{k} \, , \,\,
  L_{mo}^{[k]} = \ket{m}_{k} \bra{o}_{k} \, ,
\end{eqnarray}
for $k = 2, \ldots, N - 1$, and $L_{xw}^{[1]}$, $L_{uv}^{[1]}$,
$L_{xv}^{[N-1]}$, and $L_{uw}^{[N-1]}$ for the boundary sites. As
mentioned above, each operation in the bulk of the system introduces
an energy difference of $ \pm 4$, while at the boundaries the energy
change will only be $\pm 2$.

% -----------------------------------------------------------------------------
\subsection{Neighborhood single-channel operators in between the limits}
% -----------------------------------------------------------------------------

We explored as well the possibility of using the neighborhood single-channel
Lindblad operators from both limits to capture the region between the limits
with $0 < \phi < \pi / 2$.
An optimization over the ratio of the two coupling strengths, i.e., the
coupling strength for the neighborhood single-channel Lindblad operators
from the ferromagnetic limit and the local single-channel
Lindblad operators for the paramagnetic limit, did not lead to trace distances
close to the steady state. For example, the trace distances for values of
$\phi \approx \pi / 4$ do not drop below $0.1$. This result indicates that
the multi-channel approach is required to determine thermalization in
interacting many-body systems unless non-trivial approximations 
yet to be discovered can
simplify dynamics to neighborhood single-channel Lindblad operators.

% =============================================================================
\section{Thermalization for $S_{k} = \sigma_{k}^{z}$                            \label{app:tqi:sz}}
% =============================================================================

Up to now, all the numerical calculations in the paper have considered the
system coupling operator as $S_{k} = \sigma_{k}^{x}$. However, in this
appendix, we analyze the case $S_{k} = \sigma_{k}^{z}$. We have some
obvious changes associated with this choice. (i)~The paramagnetic limit
can now thermalize as the
eigenstates of the paramagnetic limit do not correspond to the eigenstates
of $S_{k}$. In contrast, thermalization in the ferromagnetic limit suffers
from the same effect as before in which the
Lindblad operators induce dephasing, but no
transitions. (ii)~The $\mathbb{Z}_{2}$ symmetry is broken by the Lindblad
operator, and there is no distinction between the odd and the even sector
for the choice of $S_{k} = \sigma_{k}^{z}$.

\begin{figure}[t]
  \begin{center}
    \vspace{0.8cm}
    \begin{minipage}{0.33\linewidth}
      % /data/FinalSpinsInEM3D/21_Thermal_Scales
      \begin{overpic}[width=0.95 \columnwidth,unit=1mm]{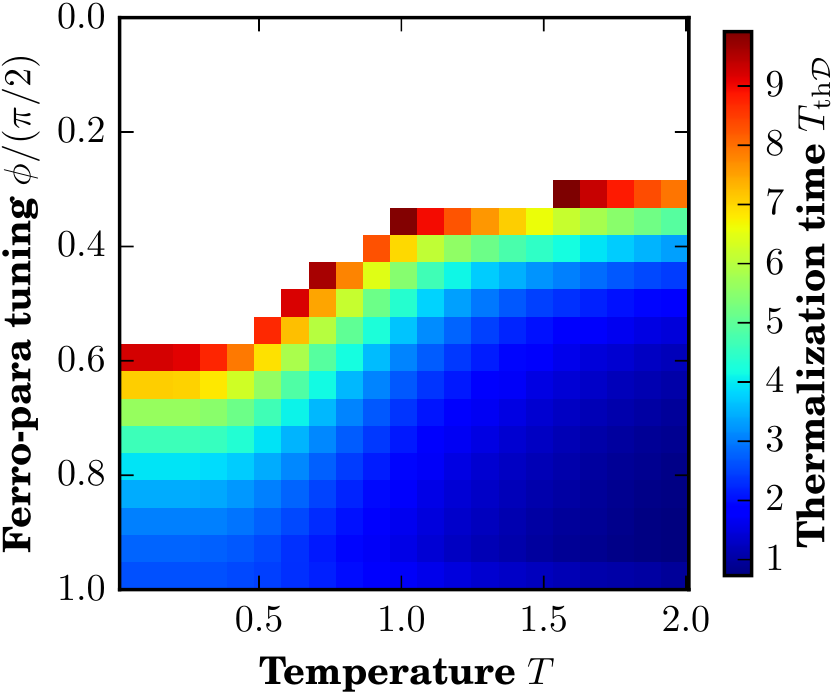}
        \put( 0,86){(a)}
      \end{overpic}
    \end{minipage}\hfill
    \begin{minipage}{0.33\linewidth}
      \begin{overpic}[width=0.95 \columnwidth,unit=1mm]{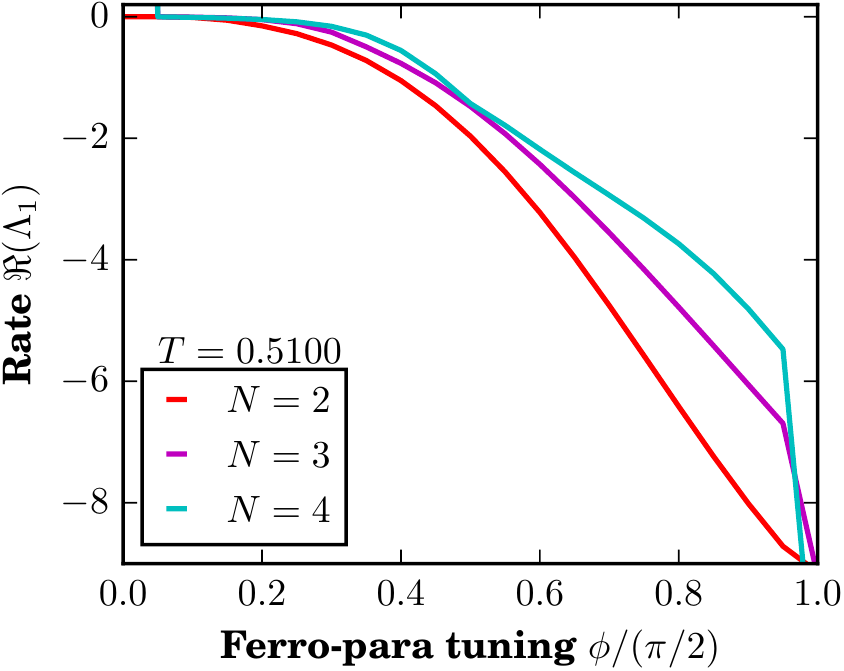}
        \put( 0,85){(b)}
      \end{overpic}
    \end{minipage}%\vspace{0.4cm}
    \begin{minipage}{0.33\linewidth}
      % /data/SpinsInEM3D/ 05_QuasiLocal_Para.py
      \begin{overpic}[width=0.95 \columnwidth,unit=1mm]{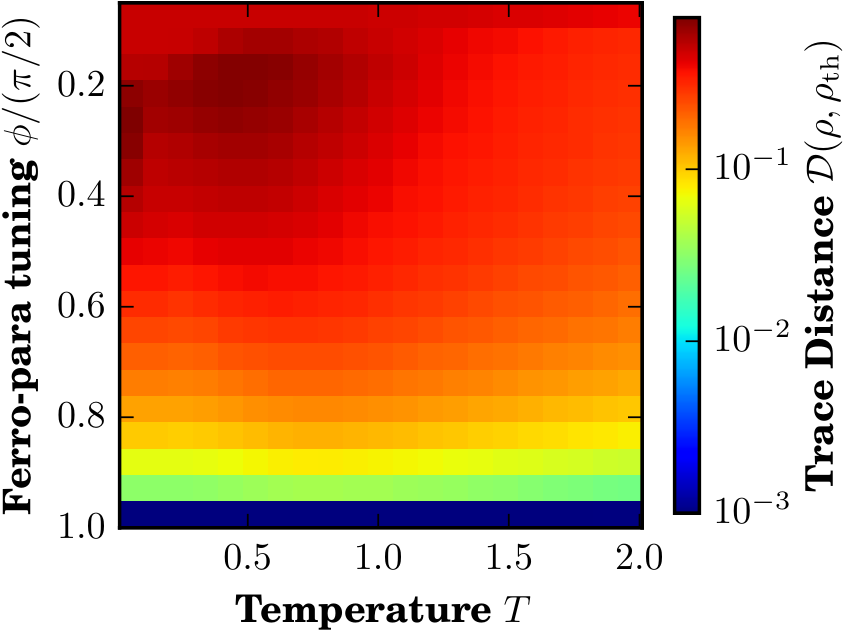}
        \put( 0,83){(c)}
      \end{overpic}
    \end{minipage}
    \caption[Summary for operator $S_{k} = \sigma_{k}^{z}$ in the interaction
      Hamiltonian]
      {\emph{Summary for operator $S_{k} = \sigma_{k}^{z}$ in the interaction
        Hamiltonian.}
      (a)~The thermalization time $T_{\mathrm{th}\mathcal{D}}$ determined
      via the trace distance exhibits longer timescales toward the ferromagnetic
      limit. White areas indicate thermalization times beyond $\tau  = 10$.
      (b)~The rate of the Liouville operator for the multi-channel approach
      with a common bath with growing system size. The rate decreases
      toward the ferromagnetic limit and decreases for increasing system sizes.
      (c)~We choose the neighborhood single-channel Lindblad operators
      designed for the paramagnetic limit on a system of $L = 4$. We observe
      that the system only thermalizes in the limit $\phi = \pi / 2$ and these operators
      cannot be used apart from $\phi = \pi / 2$. Here, we consider the final
      state of the time evolution with $\tau  = 10$ starting from a random
      initial state. The rate has non-zero values around $\phi \approx \pi / 2$
      and does not prevent thermalization.
                                                                                \label{fig:tqi:pmlimit}}
  \end{center}
\end{figure}

We present in Fig.~\ref{fig:tqi:pmlimit} the data corresponding to our
main scenario \srcasea{} and consider the multi-channel
Lindblad operators. Figure~\ref{fig:tqi:pmlimit}(a)
shows the thermalization timescale $T_{\mathrm{th}\mathcal{D}}$ for the
multi-channel approach, as defined in Eq.~\eqref{eq:tqi:msxtrd}. We observe
the opposite behavior from before, as expected: the ferromagnetic limit now
has a slow timescale indicated by the white regions where the criterion
is not met within $\tau = 10$. We study the gap as a function of the system
size in Fig.~\ref{fig:tqi:pmlimit}(b), where the gap of the Liouville
operator decreases for larger system sizes comparing $N = 2, 3$ and $4$.
Thus, analogous to Figure~\ref{fig:tqi:spectra_even}(a), we find that
the thermalization timescale also slows down for larger
systems in this case.

We now consider the single channel Lindblad operators derived in
Sec.~\ref{sec:tqi:limits} for the paramagnetic limit, which are
given by Eq.~\eqref{eq:tqi:paracool} for cooling the system, while its
Hermitian conjugate heats it. To rephrase the operators in terms of
the transition between states according to Eq.~\eqref{eq:tqi:lindblad},
we would define the states $\ket{y} = \statey{k}$ and $\ket{z} = \statez{k}$
and the corresponding Lindblad operators $L_{yz}^{[k]}$ and $L_{zy}^{[k]}$.
The corresponding energy difference used to calculate the rates is
$\Delta = \pm 2g$. These Lindblad operators correspond to spin raising
and lowering operators in the $x$-direction, and therefore they are
similar to the single-channel approaches usually considered in the
literature \cite{Cai2013,Daley2014,Rota2017,Bernier2018}.
Figure~\ref{fig:tqi:pmlimit}(c) shows that indeed the chosen single-channel
operators thermalize the system in the paramagnetic
limit. The distance to the thermal state in the limit $\phi = \pi / 2$
is below $10^{-12}$, but we set the minimal trace distance artificially to
$10^{-3}$ for better visualization of the gradient around the limit.

\end{document}